\documentclass[a4paper,fleqn,usenatbib]{mnras}

\usepackage{graphicx}	
\usepackage{amsmath}	
\usepackage{amssymb}	
\usepackage{multicol}   
\usepackage[utf8]{inputenc}
\usepackage{fancyref}
\usepackage{fullpage}
\usepackage{natbib,url,twoopt}
\usepackage{float}
\usepackage{subfloat}
\usepackage{subfig}
\usepackage{footmisc}

\title[Short title, max. 45 characters]{New Suns in the Cosmos II: Differential rotation in \textit{Kepler} Sun-like stars}

\author[M. L. Das Chagas et al.]{M. L. Das Chagas$^{1,}$$^{2}$, J. P. Bravo$^{1}$, A. D. Costa$^{1}$, C. E. Ferreira Lopes$^{6}$,\and R. Silva Sobrinho$^{1}$, F. Paz-Chinch\'on$^{7,}$$^{8}$, I. C. Le\~ao$^{3}$, A. Valio$^{5}$, D. B. de Freitas$^{1}$,\and B. L. Canto Martins$^{1}$, A. F. Lanza$^{4}$ and J. R. De Medeiros$^{1}$ \\
$^{1}$Departamento de F\'isica Te\'orica e Experimental, Universidade Federal do Rio Grande do
Norte, Natal, RN 59078-970, Brazil \\
$^{2}$Faculdade de Física - Instituto de Ciências Exatas, Universidade Federal do Sul e Sudeste do Pará, Marabá, PA 68505-080, Brazil \\
$^{3}$European Southern Observatory, Karl-Schwarzschild-Strasse 2, Garching 85748, Germany \\
$^{4}$INAF, Osservatorio Astrofisico di Catania, via S. Sofia, 78, Catania 95123, Italy \\
$^{5}$CRAAM, Universidade Presbiteriana Mackenzie, Rua da Consolação, 896, São Paulo, SP 01302, Brazil\\
$^{6}$SUPA Wide-Field Astronomy Unit, Institute for Astronomy, School of Physics and Astronomy, University of Edinburgh, \\Royal Observatory, Blackford Hill, Edinburgh EH9 3HJ, UK \\
$^{7}$Departamento de Enseñanza de las Ciencias Básicas, Universidad Católica del Norte, Larrondo 1281, Coquimbo 17814-21, Chile \\
$^{8}$Millennium Institute of Astrophysics, Av. Vicuña Mackenna 4860, Macul, Santiago 78204-36, Chile \\
}

\date{Accepted 2016 August 10. Received 2016 August 8; in original form 2016 April 5}

\pubyear{2015}

\begin{document}
\label{firstpage}
\pagerange{\pageref{firstpage}--\pageref{lastpage}}
\maketitle

\begin{abstract}
The present study reports the discovery of Sun-like stars, namely main-sequence stars with $T_{\rm eff}$, $\log g$ and rotation periods $P_{rot}$ similar to solar values, presenting evidence of surface differential rotation. An autocorrelation of the time series was used to select stars presenting photometric signal stability from a sample of 881 stars with light curves collected by the \textit{Kepler} space-borne telescope, in which we have identified 17 stars with stable signals. A simple two-spot model together with a Bayesian information criterion were applied to these stars in the search for indications of differential rotation; in addition, for all 17 stars, it was possible to compute the spot rotation period $P$, the mean values of the individual spot rotation periods and their respective colatitudes, and the relative amplitude of the differential rotation.
\end{abstract}

\begin{keywords}
stars: rotation, (stars:) starspots, stars: solar-type
\end{keywords}

\section{Introduction}
Stars are normally born with rapid rotation, and the angular velocity distribution will be established  in their infancy stages, mostly as the result of the interaction of the stellar magnetic field with the circumstellar accretion disk, at least for late-type stars (e.g., \citealt{Shu1994}; \citealt{Bouvier1997}; \citealt{2013vanSaders}). During the initial stages, the surfaces of stars with convective envelopes will slow down via magnetic braking resulting from the interaction between the stellar magnetic field and the magnetized wind from the surface (e.g., \citealt{1988Kawaler}; \citealt{2012Reiners}). 

Surface rotation can now be measured for many families of stars using different procedures, including the analyses of spectral line broadening, which produces projected rotational velocity \textit{v} sin(\textit{i}) measurements (e.g., \citealt{1999Medeiros}; \citealt{2014Medeiros}; \citealt{2004Nordstrom}), and periodic modulation of starlight produced by non-uniformities on the surface of the stars (e.g., \citealt{2012Affer}; \citealt{2013Medeiros}; \citealt{2014McQuillan}; \citealt{2015Leao}). Other procedures include those based on line core variations in the Ca II H and K lines (e.g., \citealt{1983Baliunas}) and on the Rossiter-McLaughlin effect or ellipsoidal light variations in eclipsing binaries. 

In addition, it is now well established that the surface and internal stellar rotation pattern is by no means uniform. For instance, Helioseismology has revealed a large spread of rotation rates in the  outer convective regions at different latitudes, with the inner regions presenting an almost constant rotation rate (e.g., \citealt{2010Aerts}). These aspects are intimately associated with the stellar differential rotation (hereafter DR), i.e., the property that different parts of the star rotate at different rates (\citealt{2005Miesch}; \citealt{2009Miesch}; \citealt{2013Kitchatinov}). The current leading theoretical basis, first presented by \citet{1941Lebedinsky}, explains differential rotation based on the interaction between convection and rotation, with convective motions in a rotating star being disturbed by the Coriolis force. Its back reaction redistributes angular momentum and disturbs the global rotation behavior to produce non-uniformities, leading to DR of the surface.

Different procedures can be used in the diagnostic of surface DR. In the first procedure, Doppler imaging, the positions of individual spots are estimated based on their effects on the stellar spectral line profiles, on the condition that the star is rotating rapidly enough (e.g., \citealt{2002Collier}). In the second procedure, the 
Fourier transform method, the Doppler shift at different latitudes due to rotation can be estimated from the Fourier transform of the line 	profiles (e.g., \citealt{2003Reiners}; \citealt{2013Reiners}). In the third procedure, time series photometry, the rotation periods can be computed from a time	series of photometric observations (e.g., \citealt{2015Aigrain}; \citealt{2014Lanza}; \citealt{2015Davenport}). Another approach is based on Asteroseismology, in which the frequency splitting of global oscillations is explained in terms of different latitudinal rotation rates (e.g., \citealt{2004Gizon}). A recent blind survey of competing techniques for detecting rotation and DR from model photometry, conducted by \citet{2015Aigrain}, showed excellent agreement in recovering the overall rotation periods for stars exhibiting low and moderate activity levels. However, the referred study revealed a complex degeneracy  between DR shear, spot lifetimes, and the number of spots present, suggesting that DR studies based on full-disc light curves alone need to be treated with caution.

The advent of the space-borne CoRoT \citep{2006Baglin} and \textit{Kepler} \citep{2010Borucki} telescopes made it possible to study in great detail the behavior of the rotation of Sun-like stars. In this context, a large effort is being directed at the analysis of more active stars using the photometric modulations observed from their light curve (e.g., \citealt{2012Frohlich}; \citealt{2012Bonomo}; \citealt{2013McQuillan}; \citealt{2013Medeiros}), therein producing rotation periods for thousands of different families of stars. A parallel effort is being made by different authors to enlarge the horizons of our quantitative and qualitative understanding of DR (e.g., \citealt{2013Reinhold}; \citealt{2013Reiners}; \citealt{2014Lanza}; \citealt{2015Aigrain}; \citealt{Reinhold}).

Most of the DR surface patterns observed to date are predominantly solar-type, with rotation rates decreasing from the equatorial to polar regions (e.g., \citealt{2013Reiners}; \citealt{2014Lanza}; \citealt{2002Collier}; \citealt{2003Reiners}; \citealt{1993Lanza}; \citealt{1983Baliunas}). The DR total surface gradient varies to a high degree with the effective temperature (\citealt{2005Barnes})  and to a low degree with the rotation rate (\citealt{2005Kuker}). Antisolar DR measurements are sparse and have mainly been performed for some late-type giant stars, most of which being components of RS CVn-systems (e.g., \citealt{2003Strassmeier}; \citealt{2003Olah}; \citealt{2005Weber}; \citealt{2007Vida}). As noted by different authors (e.g., \citealt{2015Kovari}), it appears that the strength and even the orientation of the DR are influenced by close companions, although such a scenario is not yet understood.

By applying asteroseismology procedures to time series obtained from light curve (hereafter LC) data from the \textit{Kepler} or CoRoT missions, we are now in a position to extract, in addition to information about the surface rotational pattern, the physical characteristics of the stellar interior, revealing not only relevant aspects of DR but also information about pulsation modes and important constraints for dynamo models of low-mass stars. This enables one to test theoretical models for internal DR (see, e.g., \citealt{2011Kitchatinov}; \citealt{2011ANKuker}), as well as the development of 3D simulations (e.g., \citealt{2004Brun}; \citealt{2008Browning}; \citealt{2012Kapyla}). Further, it has been possible to estimate the ratio between the rotation rate in the small helium core and the large convective regions of late-type stars (e.g., \citealt{2010Eggenberger}). 

This is the 2nd paper of a series of studies devoted to the identification of Sun-like stars presenting physical properties similar to the Sun. In the 1st study \citep{2013Daniel}, we identified stars representing potentially good matches to the Sun's rotation. The main goal of the present work is to apply spot modeling \citep{2014Lanza} for a large sample of Sun-like stars observed in the scope of the \textit{Kepler} mission, therein attempting to measure DR and quantify how common DR is among Sun-like stars presenting solar parameters and, in particular, stars with similar Sun rotation periods.
\begin{figure}
	\centering 
	\includegraphics[width=0.52\textwidth]{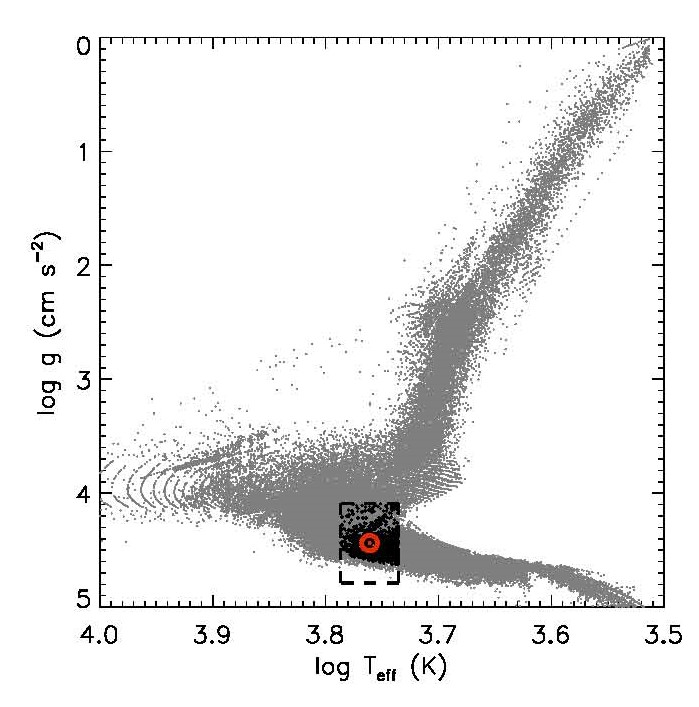} 
	\caption[]{Distribution of $\log g$ and $T_{\rm eff}$ from entire \textit{Kepler} database. The black rectangle denotes the region of sources with solar parameters with $T_{\rm eff}^{\odot}$ and $\log g^{\odot}$. The red circle shows the position of  the Sun, and the small cross  is the sample analysed.}
	\label{loggteff}
\end{figure}

\begin{figure*} 
	\centering 
	\subfloat{ \label{wvm:38}\includegraphics[angle=90, scale=0.21]{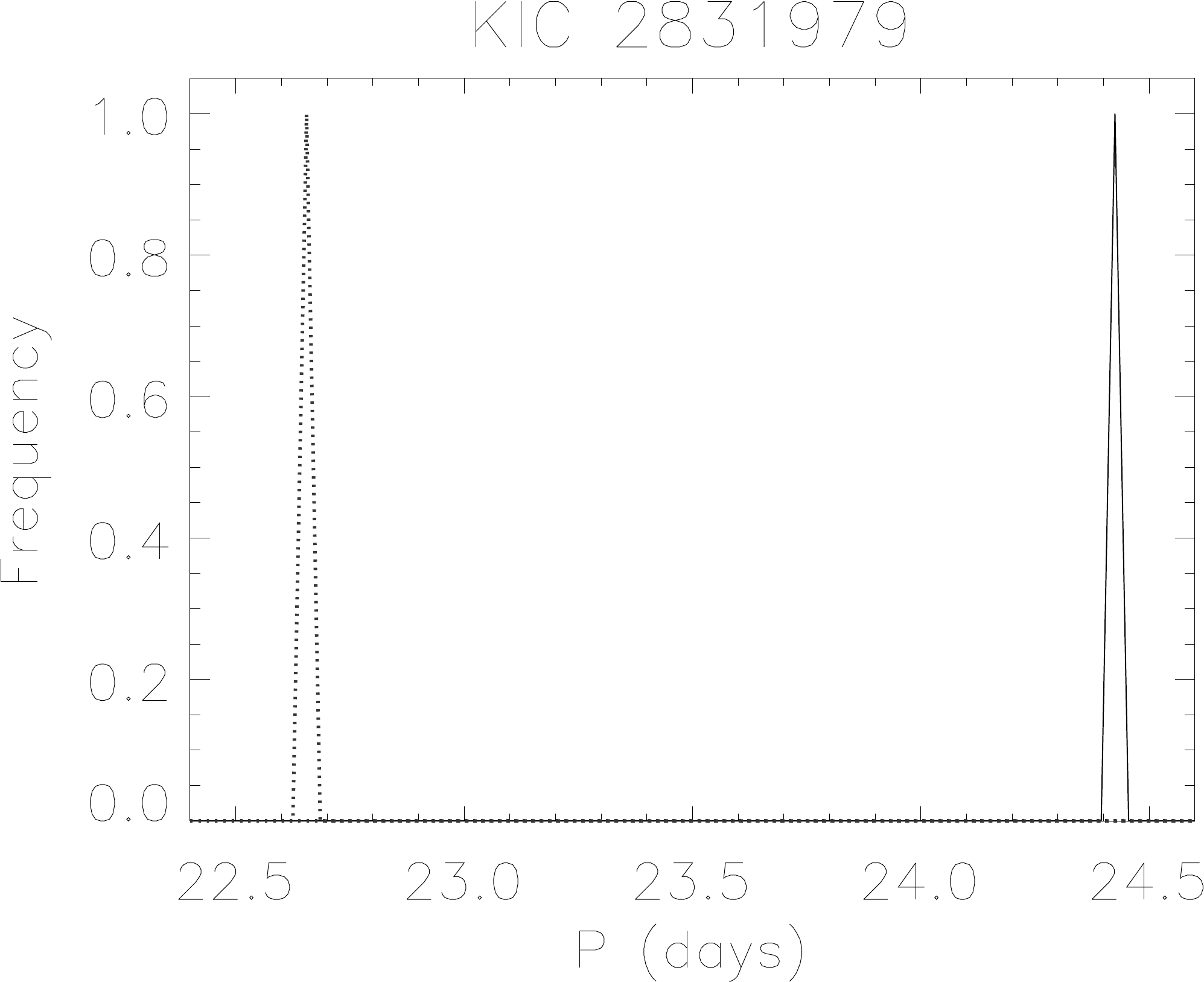} } \,
	\subfloat{ \label{wvm:40}\includegraphics[angle=90, scale=0.21]{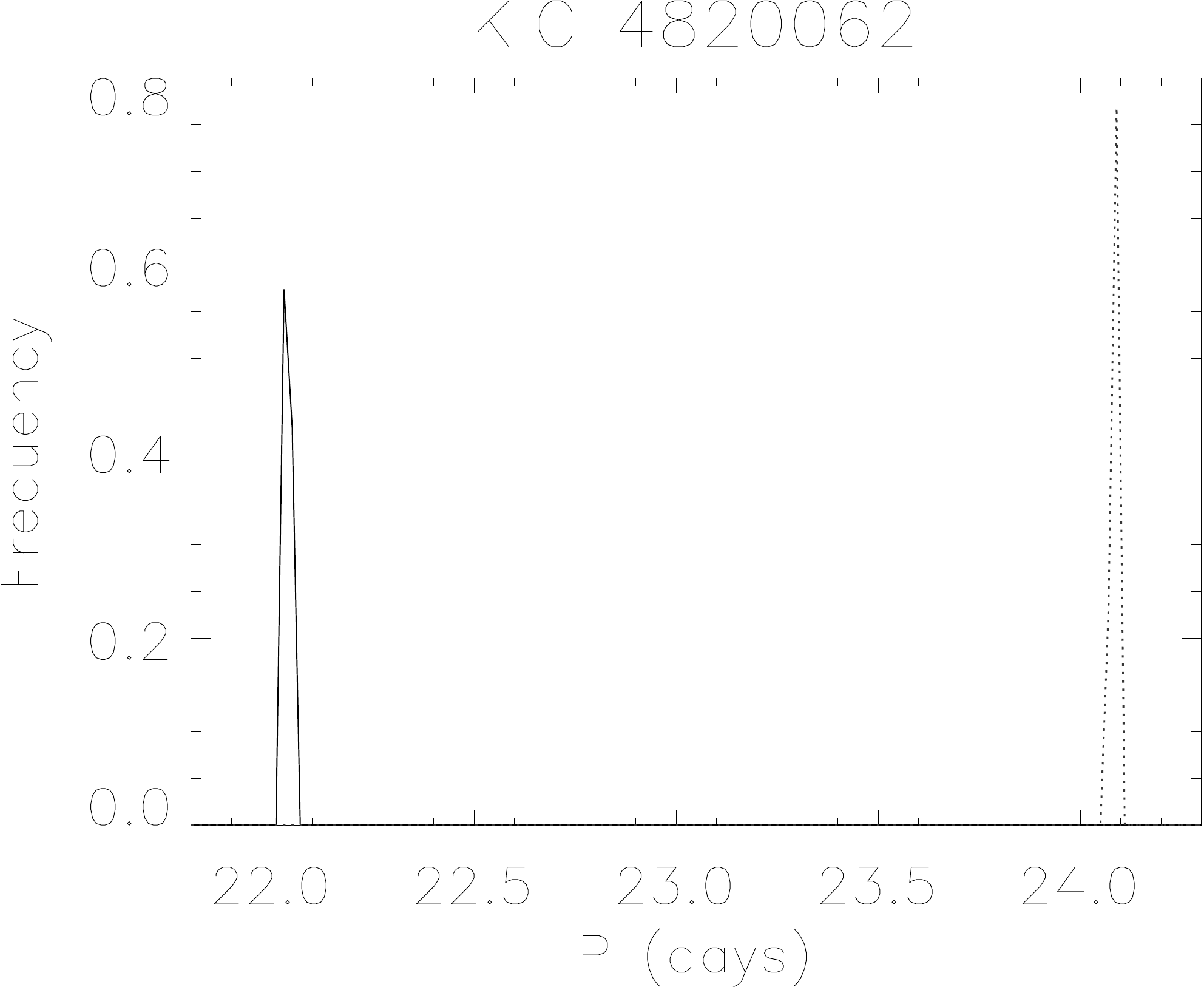} } \,
	\subfloat{ \label{wvm:41}\includegraphics[angle=90, scale=0.21]{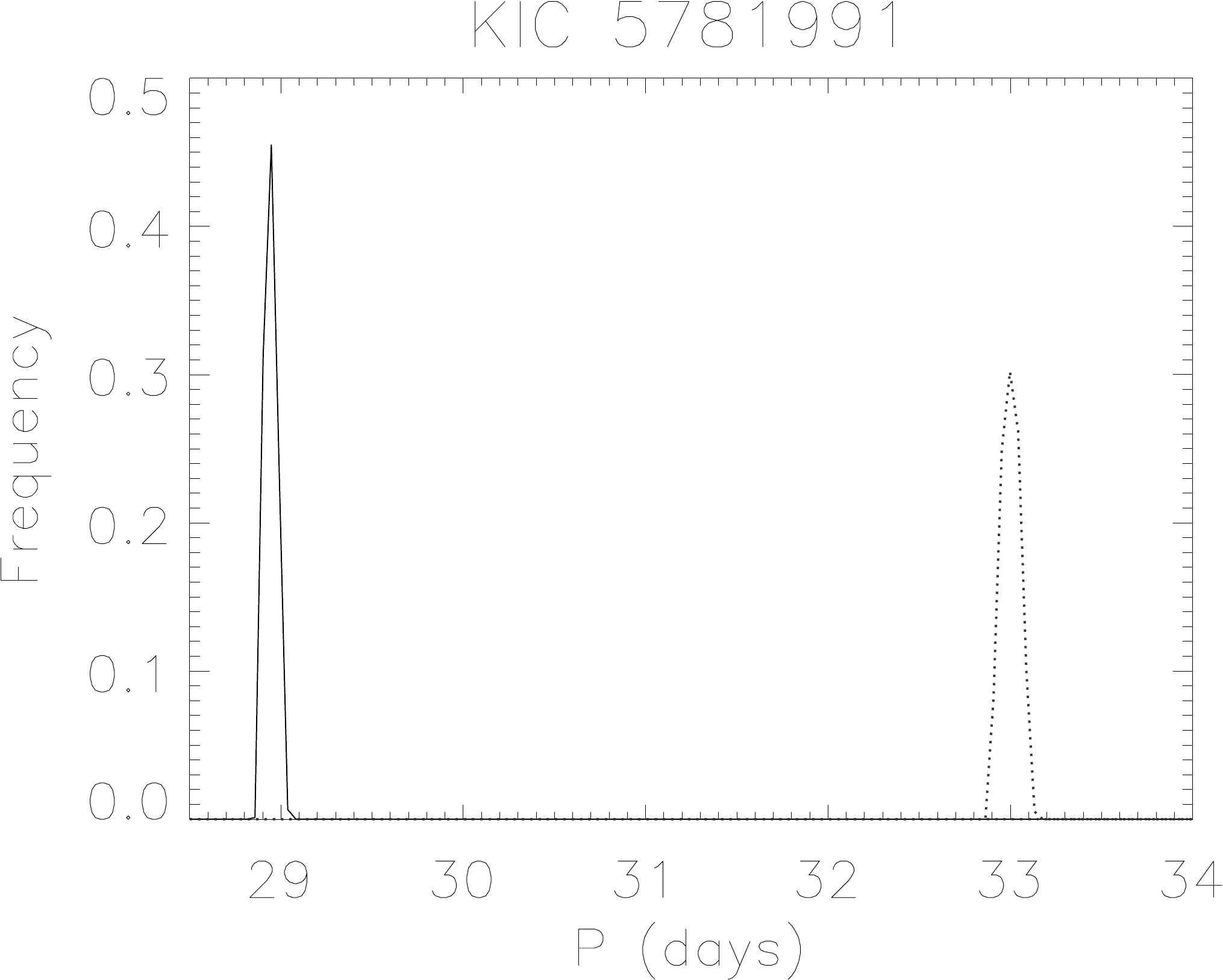} } \,
	\subfloat{ \label{wvm:42}\includegraphics[angle=90, scale=0.21]{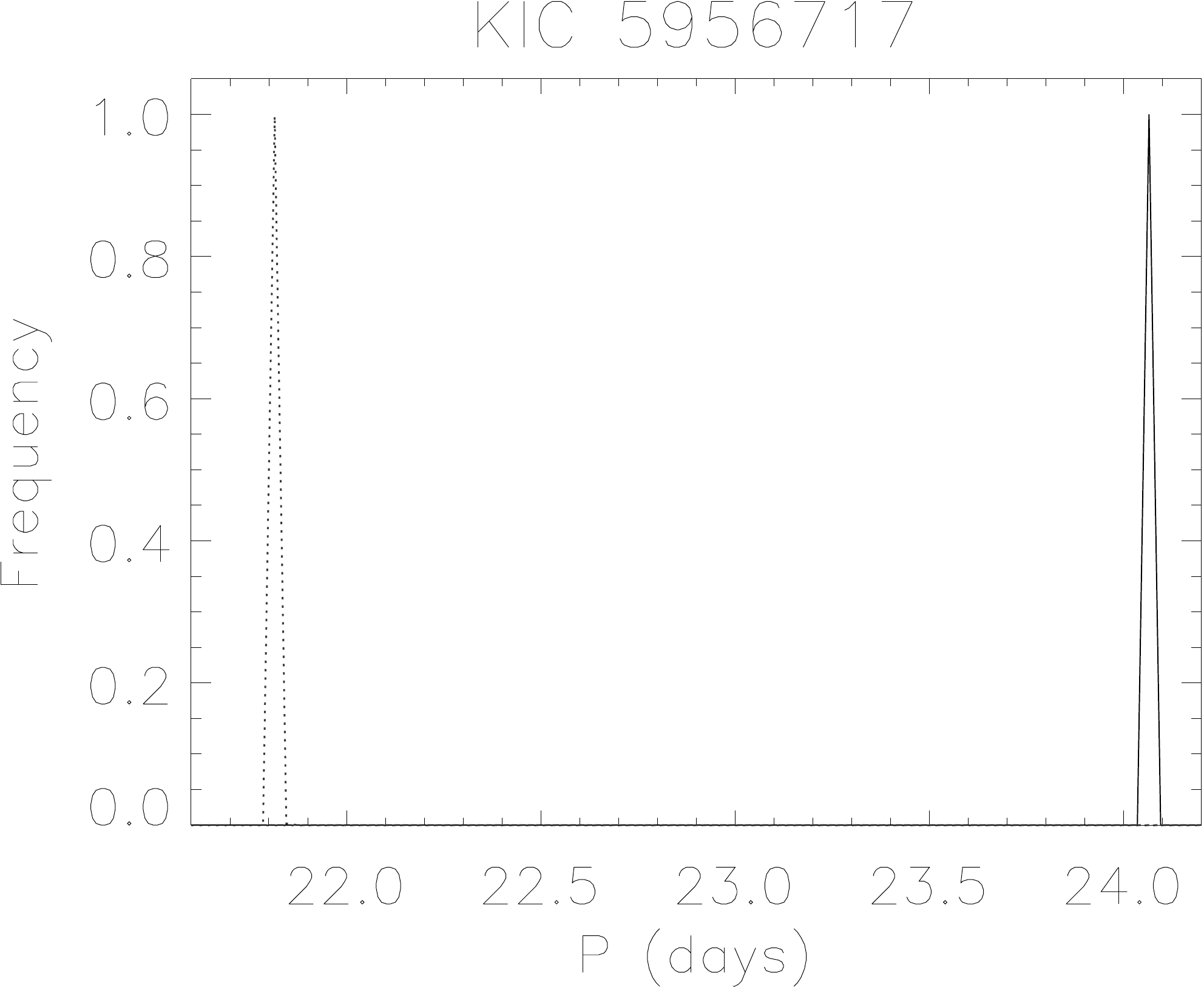} } \,
	\subfloat{ \label{wvm:43}\includegraphics[angle=90, scale=0.21]{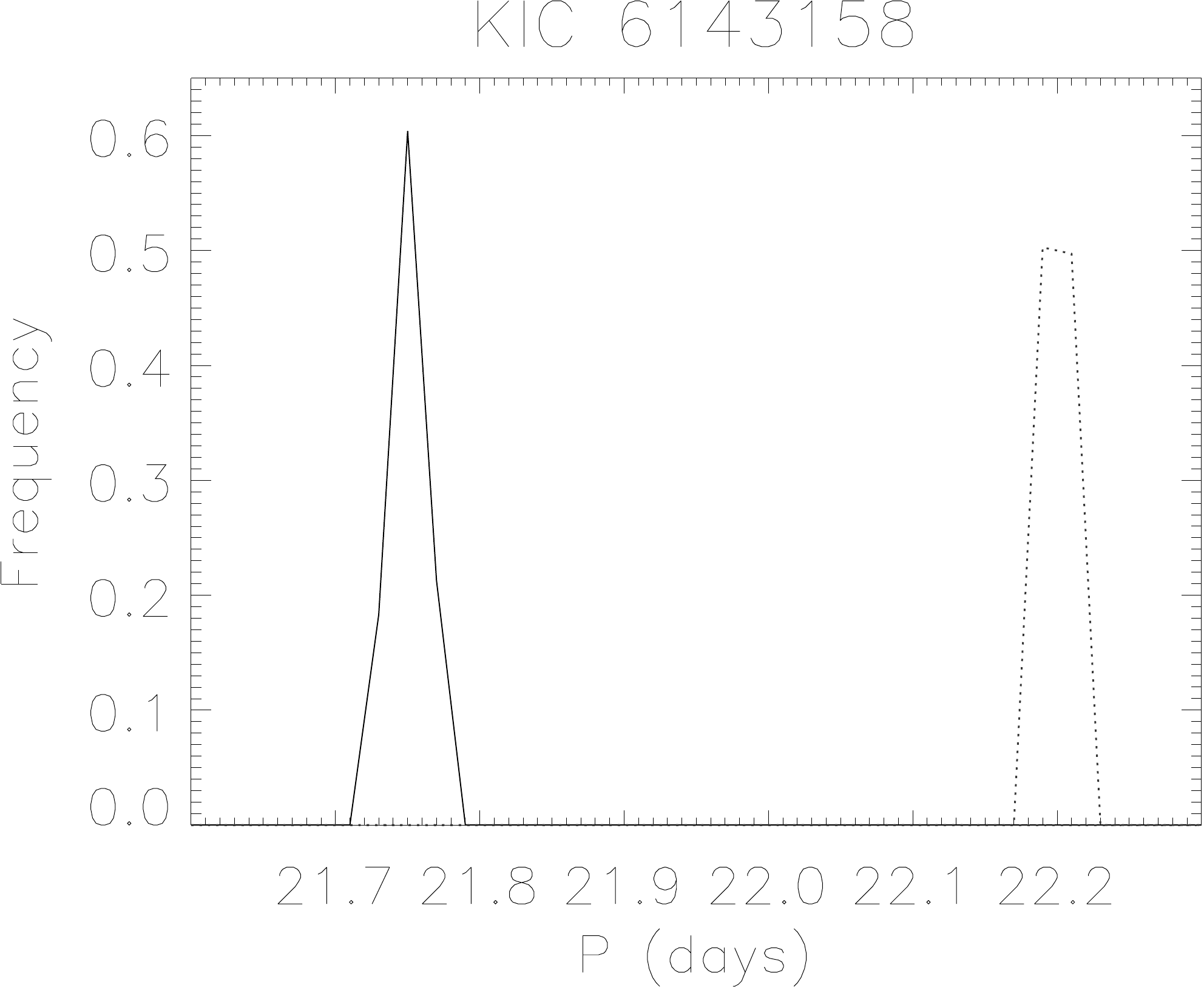} } \,
	\subfloat{ \label{wvm:44}\includegraphics[angle=90, scale=0.21]{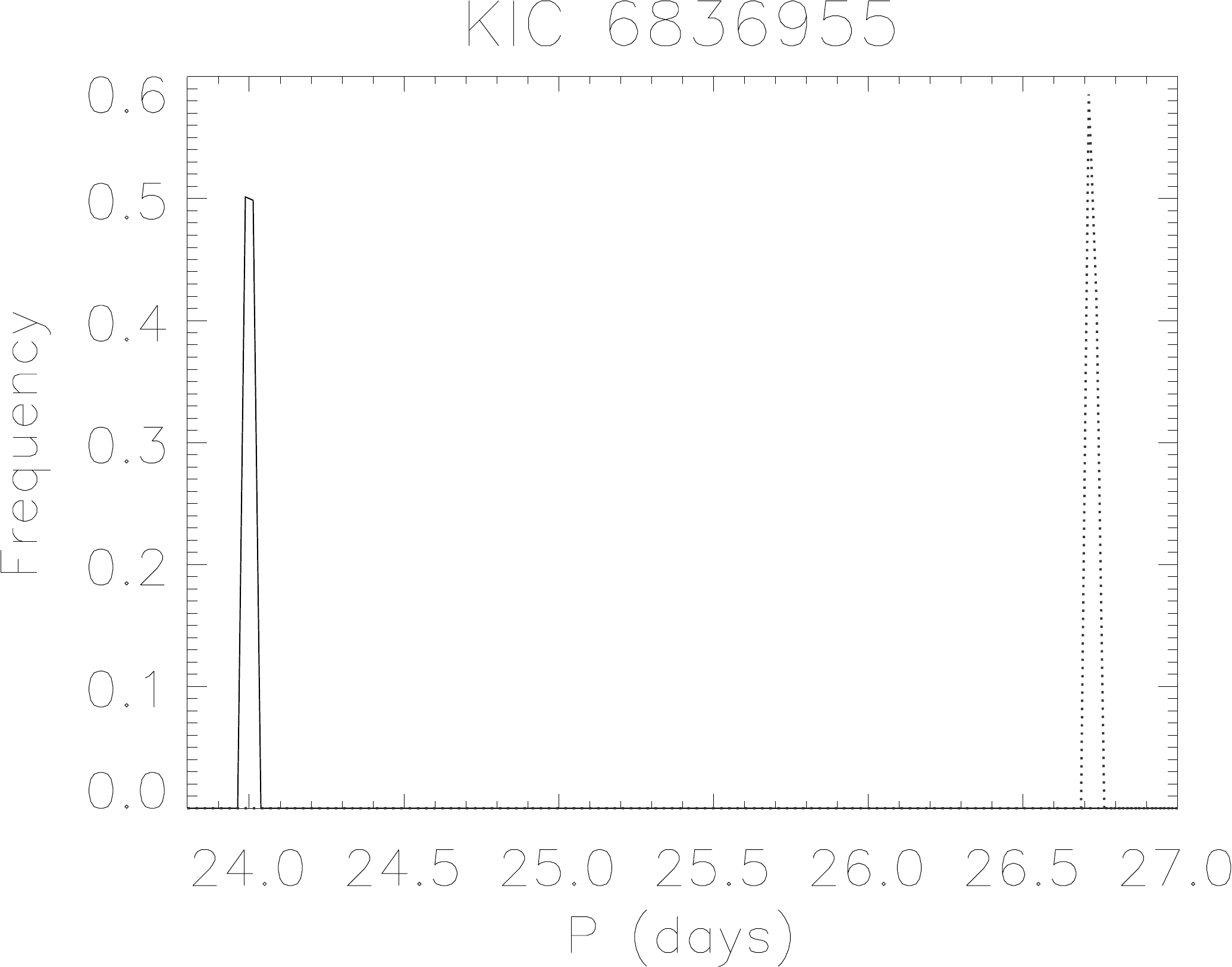} } \,
	\subfloat{ \label{wvm:45}\includegraphics[angle=90, scale=0.21]{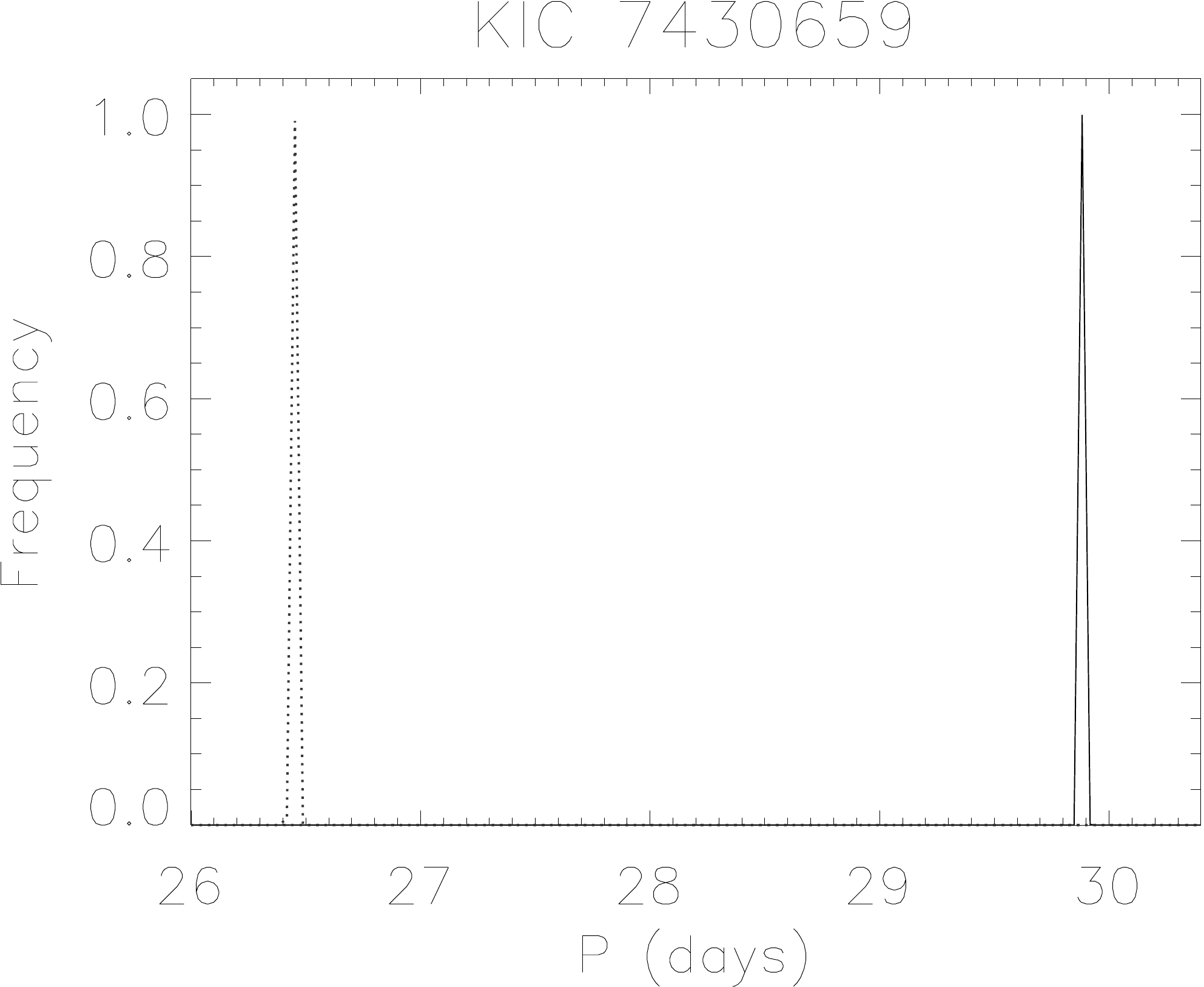} } \,
	\subfloat{ \label{wvm:46}\includegraphics[angle=90, scale=0.21]{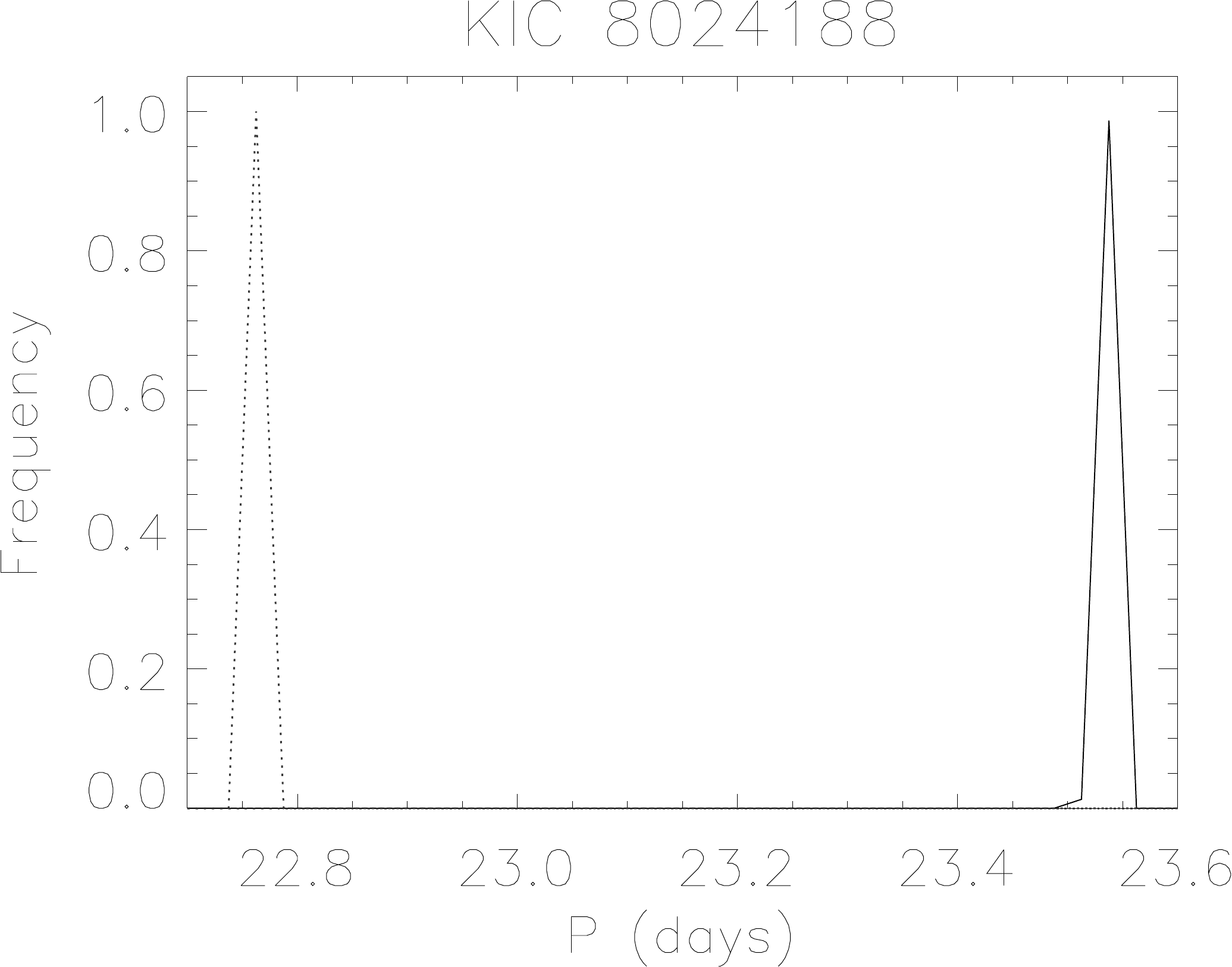} } \,   
	\subfloat{ \label{wvm:47}\includegraphics[angle=90, scale=0.21]{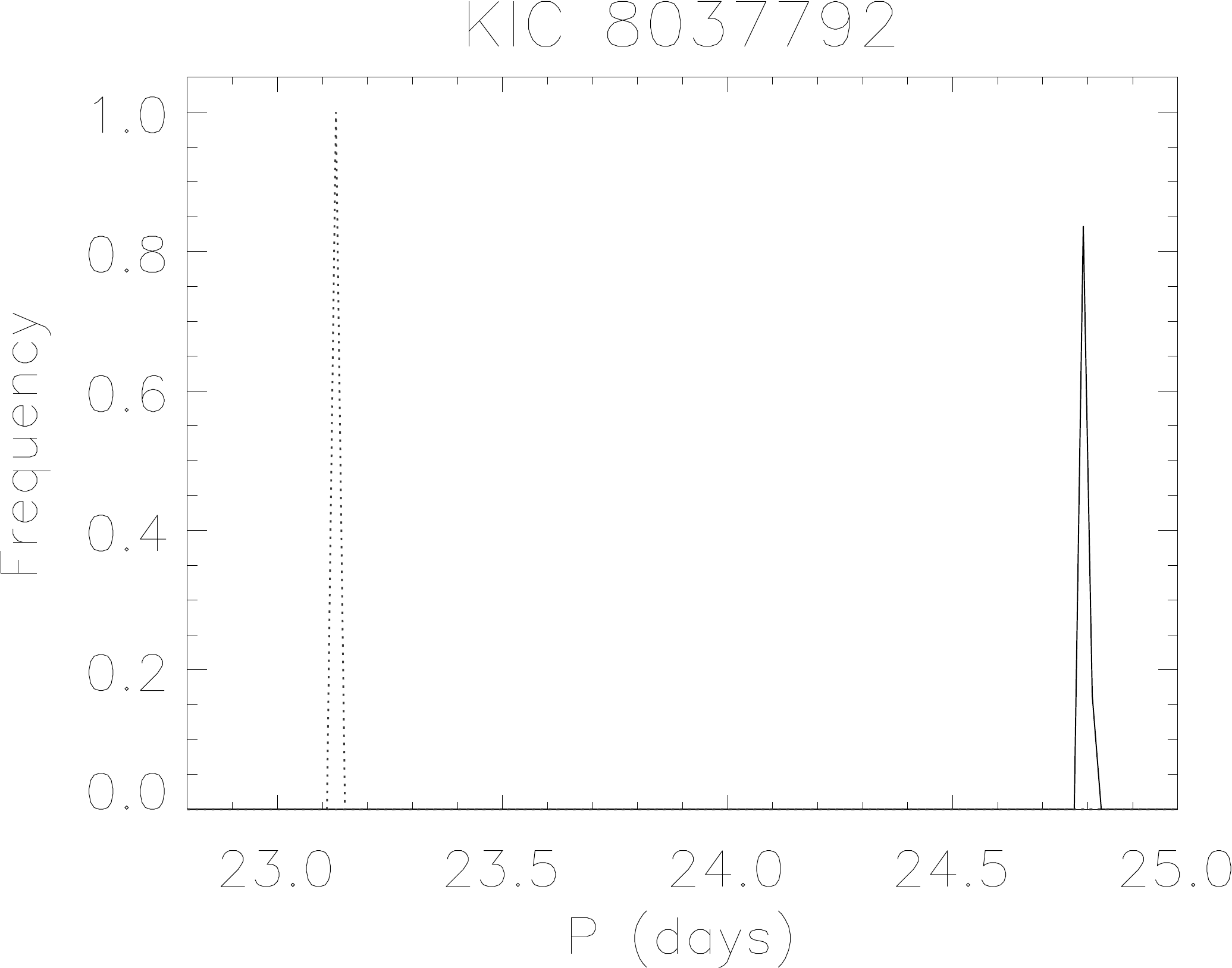} } \,
	\subfloat{ \label{wvm:48}\includegraphics[angle=90, scale=0.21]{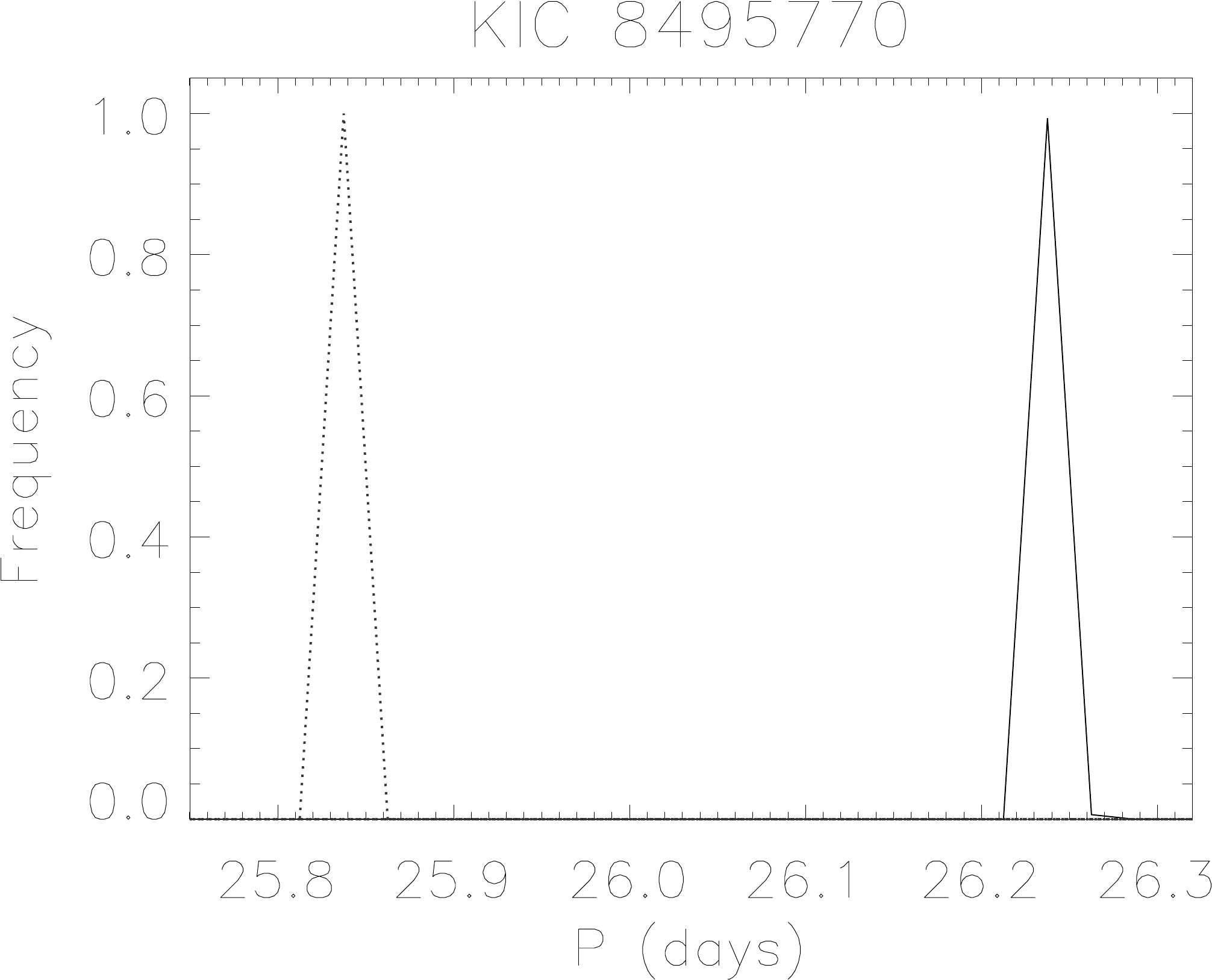} } \,
	\subfloat{ \label{wvm:49}\includegraphics[angle=90, scale=0.21]{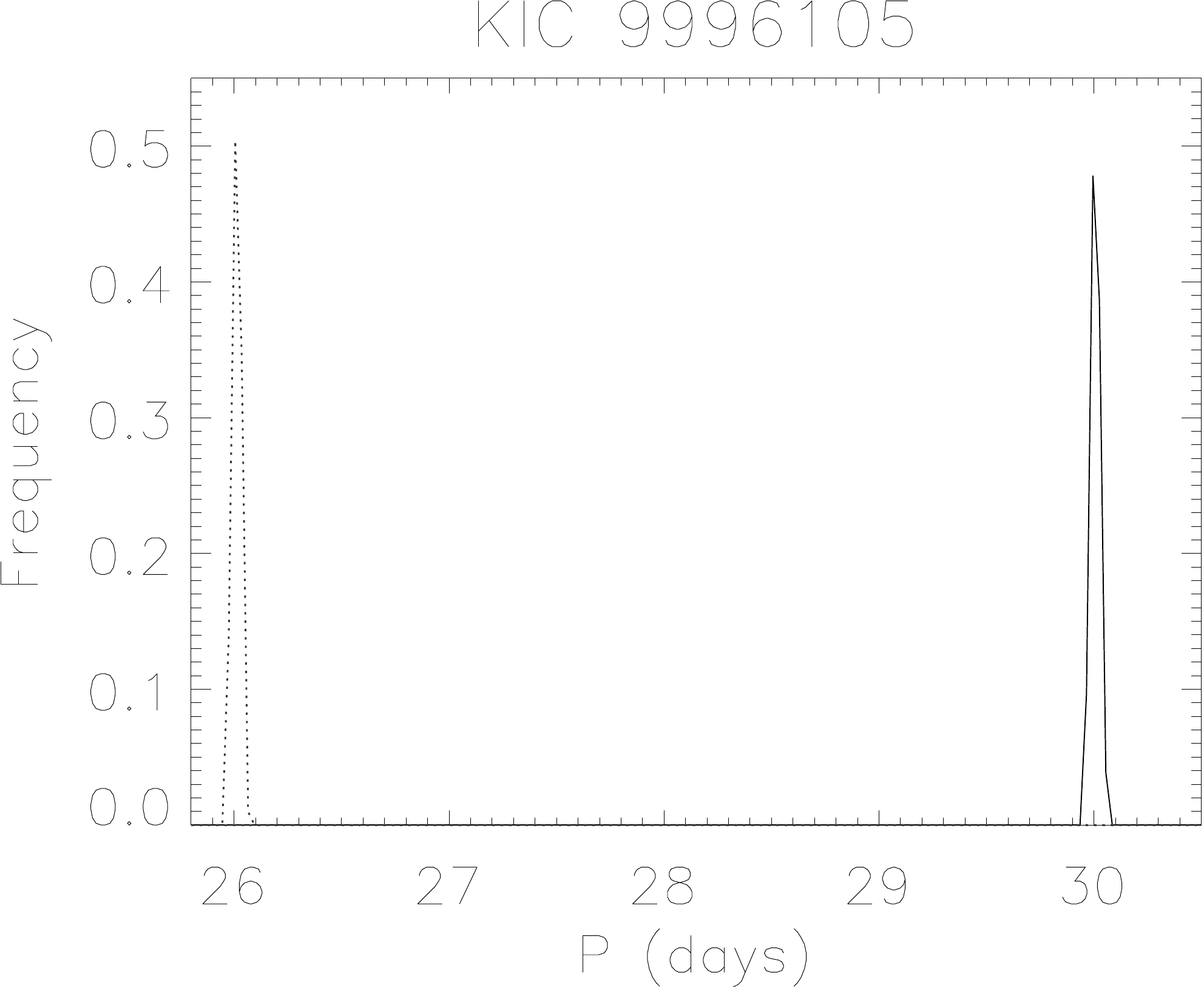} } \,
	\subfloat{ \label{wvm:50}\includegraphics[angle=90, scale=0.21]{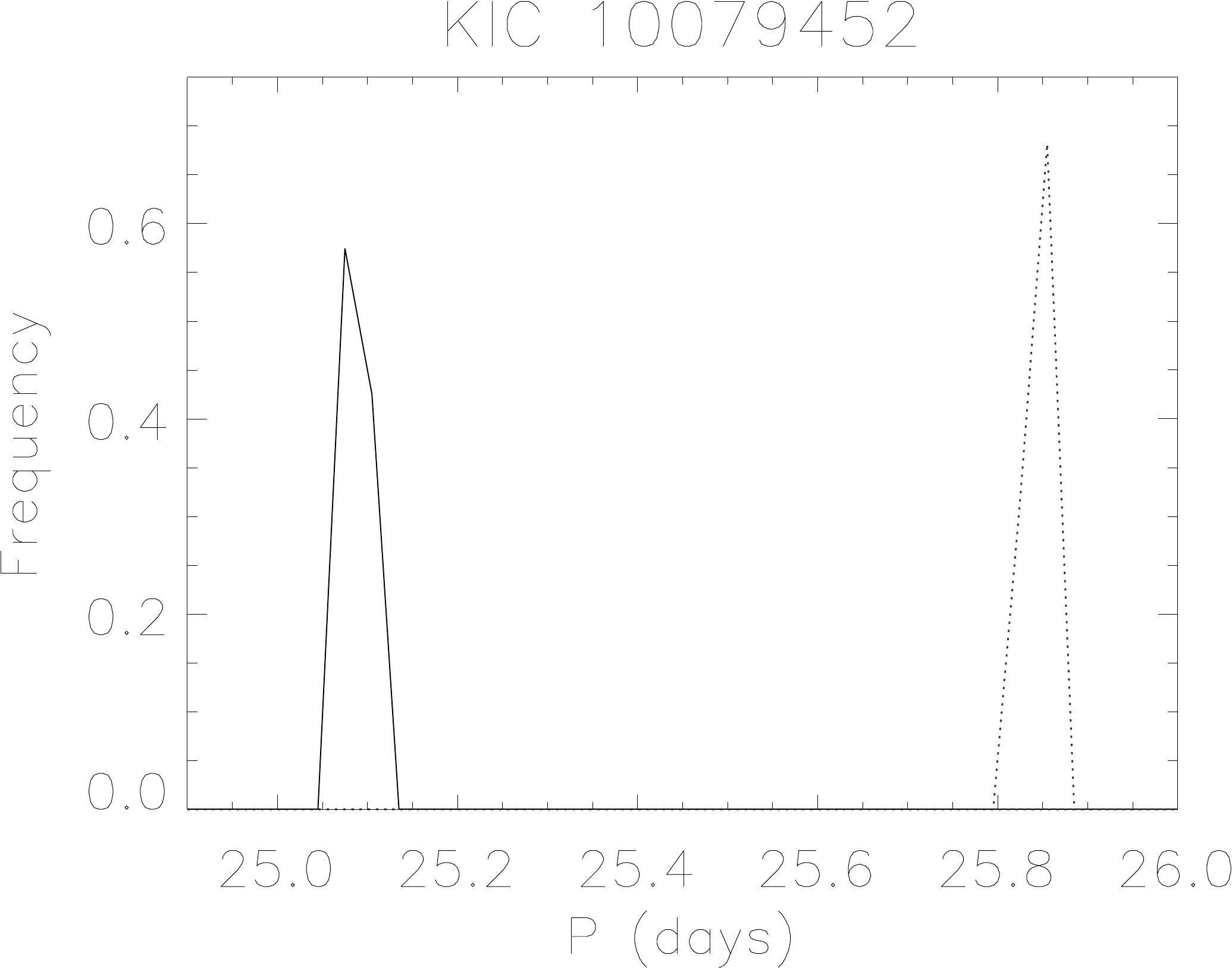} } \,
	\subfloat{ \label{wvm:51}\includegraphics[angle=90, scale=0.21]{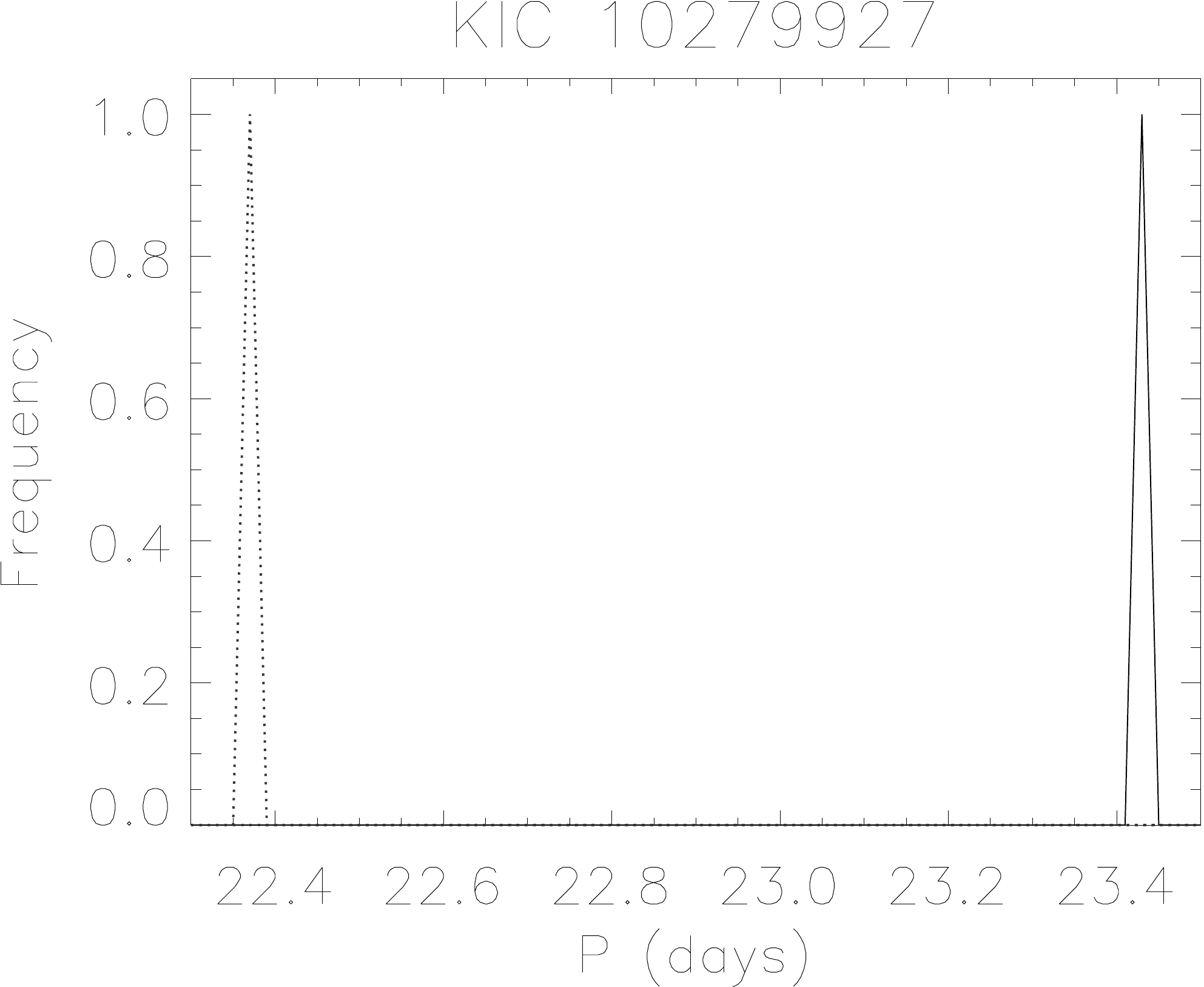} } \,
	\subfloat{ \label{wvm:52}\includegraphics[angle=90, scale=0.21]{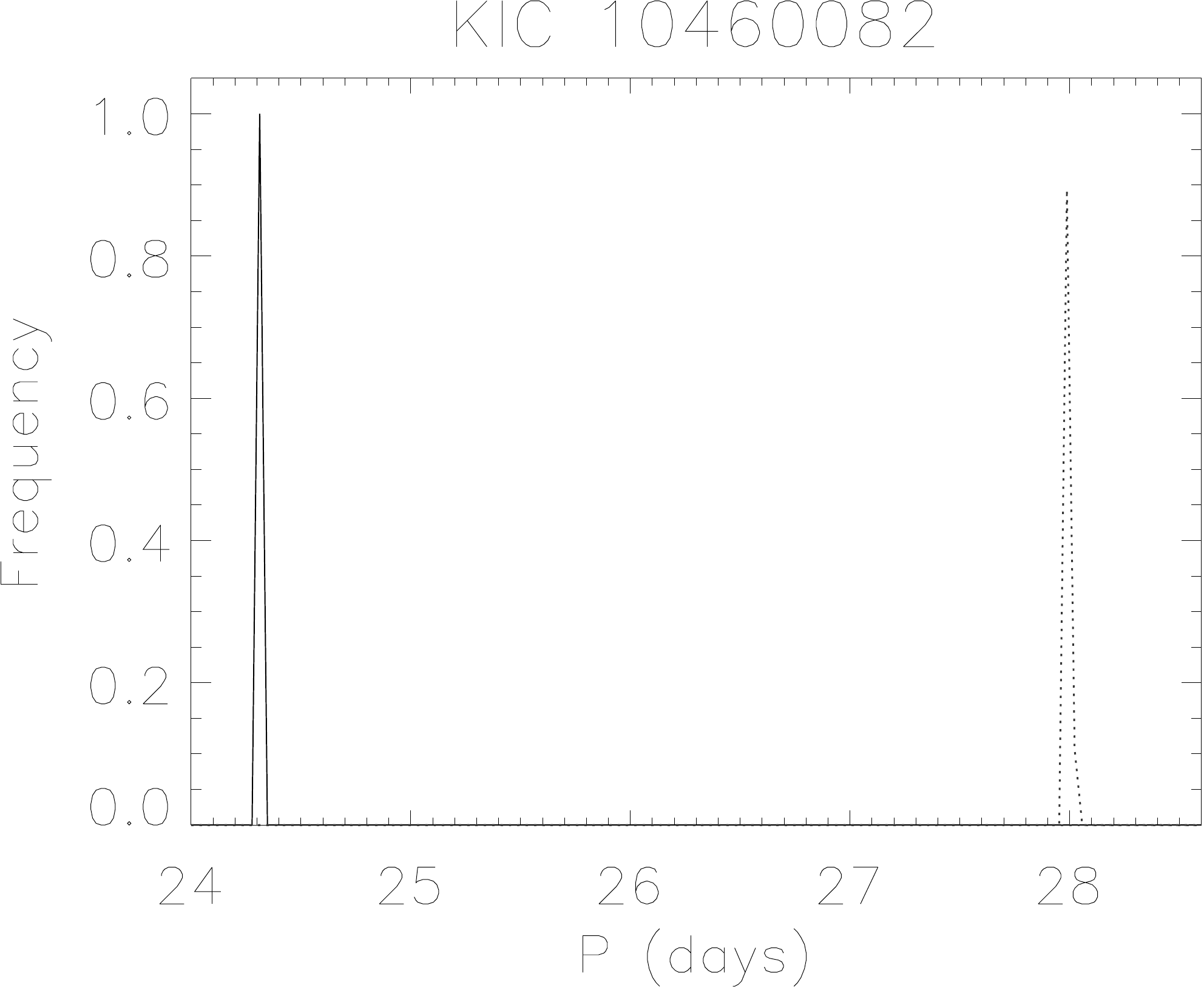} } \,
	\subfloat{ \label{wvm:53}\includegraphics[angle=90, scale=0.21]{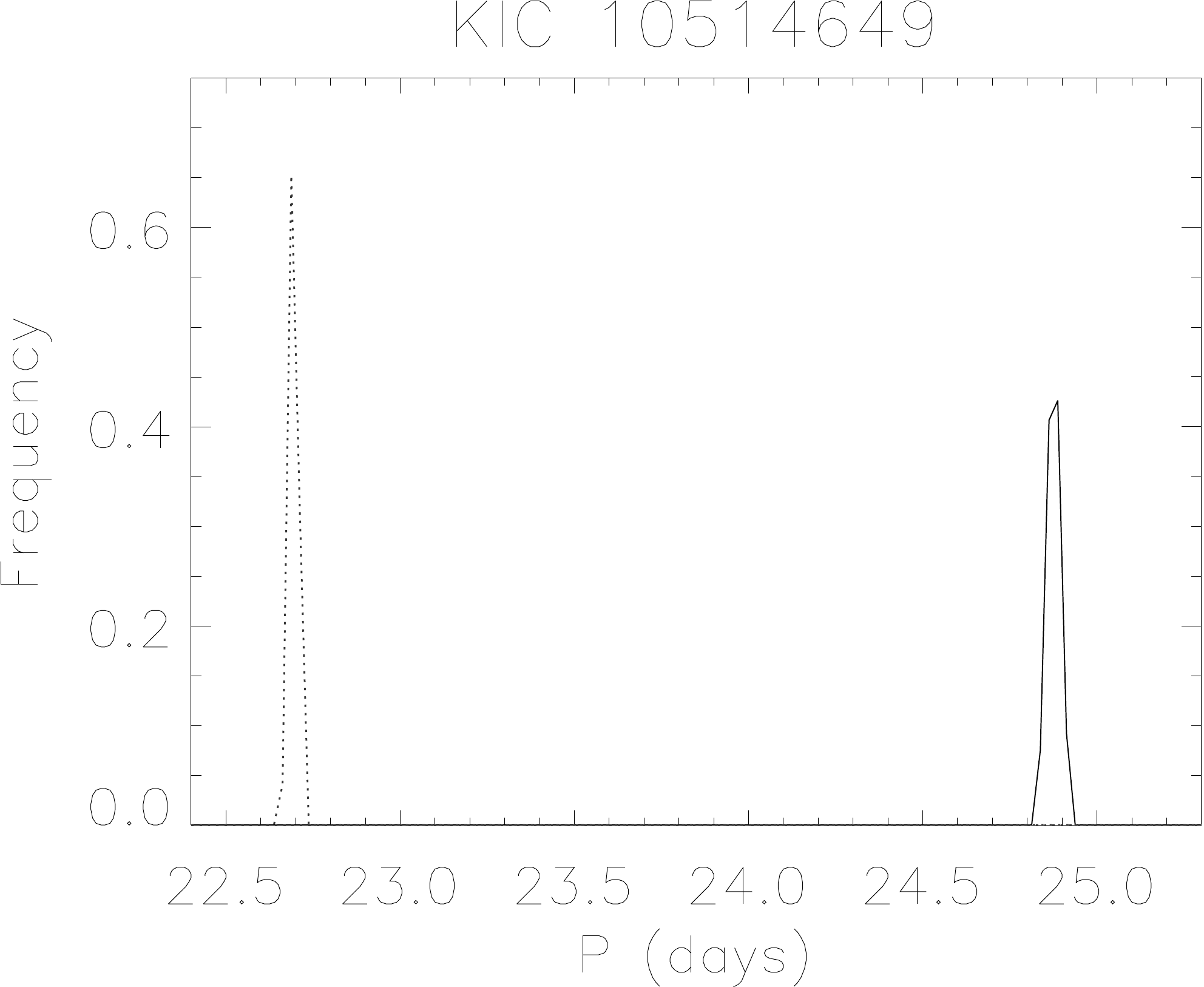} } \,
	\subfloat{ \label{wvm:54}\includegraphics[angle=90, scale=0.21]{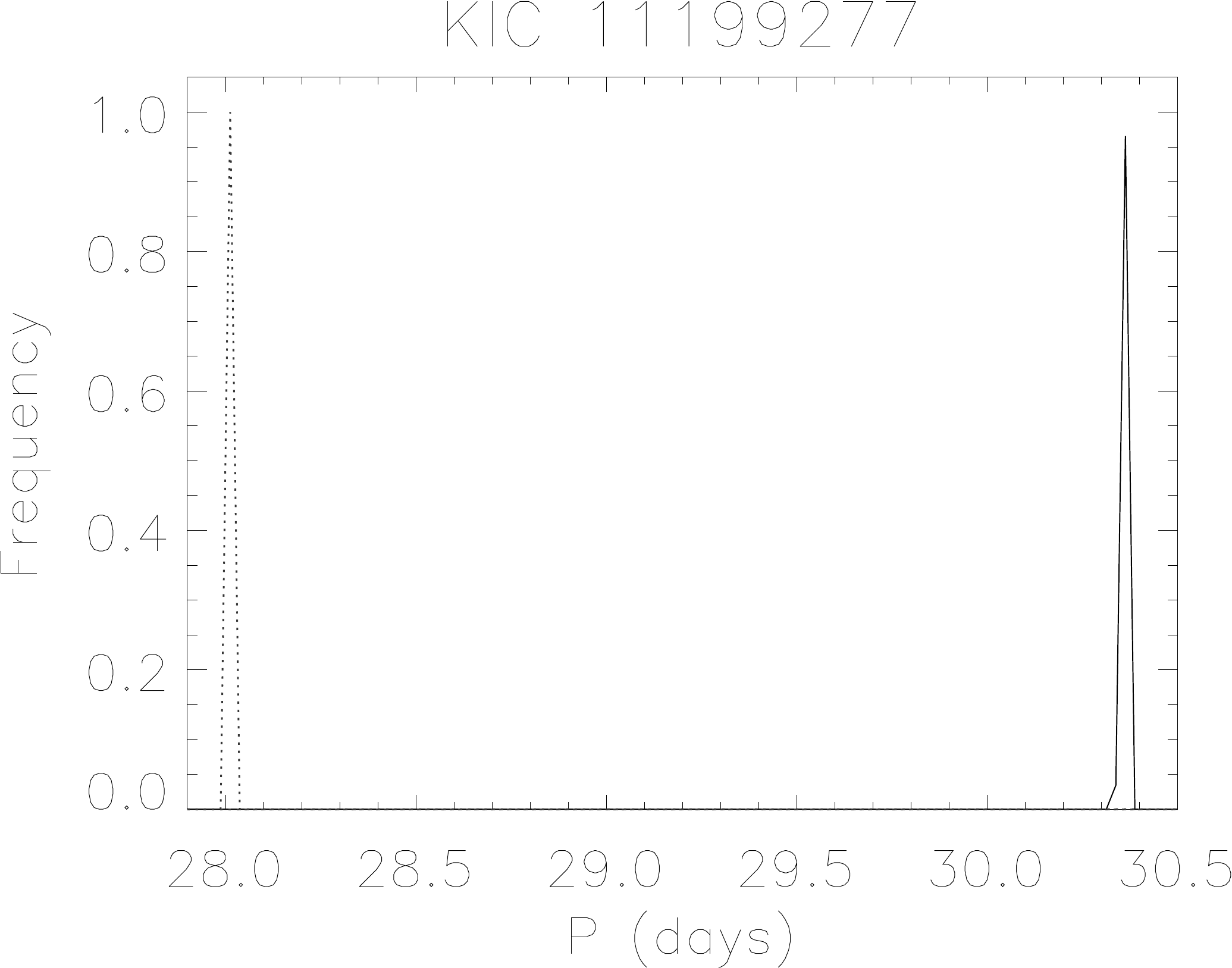} } \,
	\subfloat{ \label{wvm:55}\includegraphics[angle=90, scale=0.21]{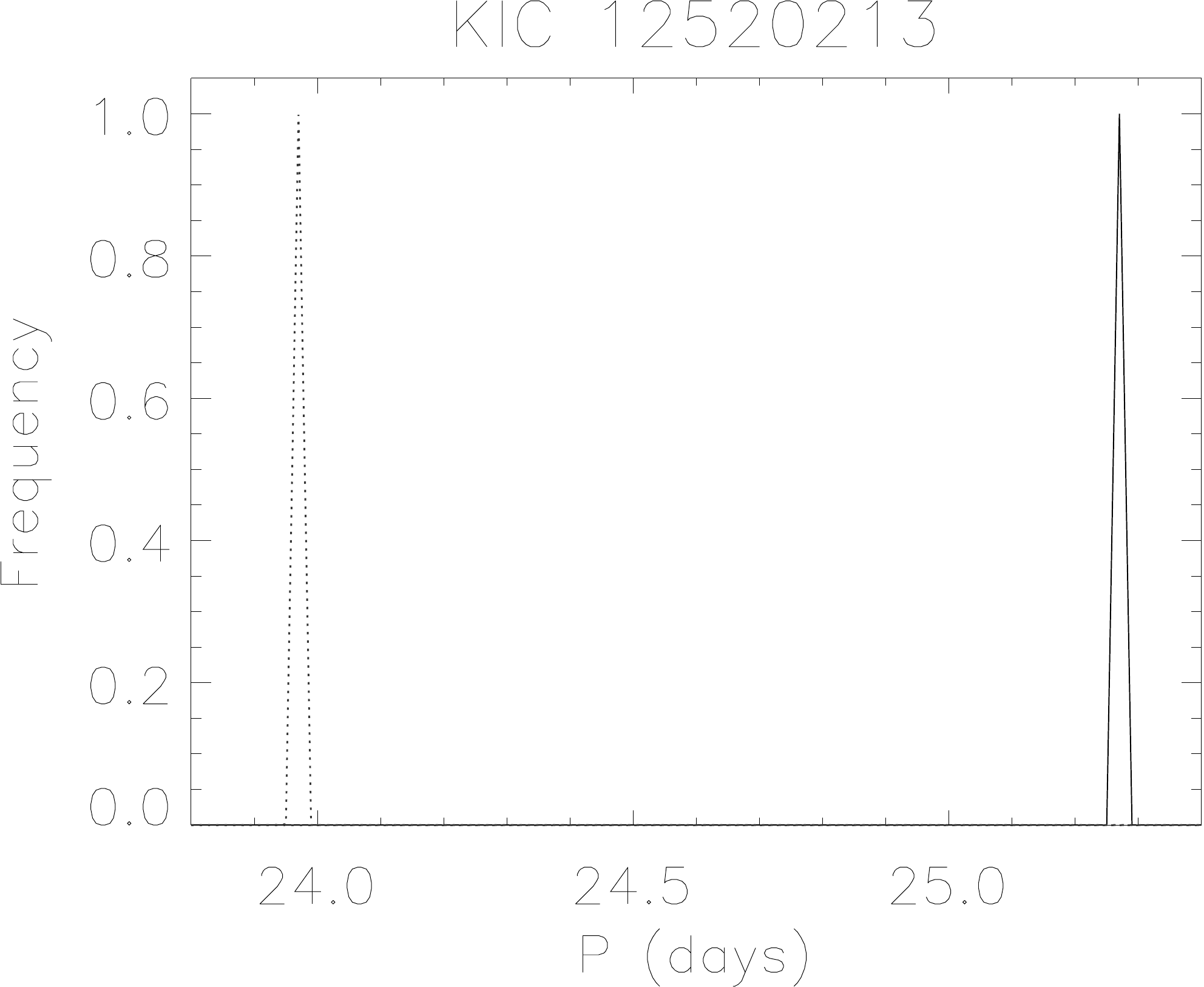} } \,	
	\caption{A posteriori distributions of the rotation periods of the two spots as derived from MCMC for all stars. The solid line refers to the distribution of the rotation period of the first spot, and the dashed line refers to that of the second.}
	\label{map}
\end{figure*}
This paper is organized as follows: Section \ref{sec:observations} presents the stellar sample with the \textit{Kepler} stellar parameters. In Section \ref{sec:autocorr}, we introduce the autocorrelation function. Results and Conclusions are presented in Sections \ref{sec:results} and \ref{sec:concl}, respectively.

\section{Working sample and data analyses} \label{sec:observations}

From May 2009 to May 2013, the \textit{Kepler} mission collected data in a steady field of view for 191,449 stars in 17 runs (known as quarters), which were composed of long-cadence (6.02 s observations stacked every 29.4 minutes, \citep{2010Jenkins}) and short-cadence (bins of 59 s) observations (\citealt{2010Van};\citealt{2013Thompson}). For the present study, we selected the calibrated LCs processed by the PDC\_MAP pipeline \citep{2010Jenkins}. To search for stars with physical properties approximately equal to the solar values, we made an initial selection of LCs from the \textit{Kepler} database (Mikulski Archive for Space Telescopes\footnote{http://archive.stsci.edu/kepler/data\_search/search.php, hereafter MAST}) using the solar parameters $\log g$ ($\sim$ 4.44) cm s$^{-2}$, $T_{\rm eff}$ ($\sim$ 5779) K, $[Fe/H]$ ($\sim$ 0.) dex  and 23 days $< P_{rot} <$ 33 days. A total of 881 stars with 3.94 cm s$^{-2} < \log g <$ 4.94 cm s$^{-2}$, 5579 K $< T_{\rm eff} <$ 5979 K,  were selected, with effective temperature and gravity obtained from \citet{2014Huber} and rotation period given by \citet{2013McQuillan}. The location of our working sample, in the $\log g$ versus $T_{\rm eff}$ diagram, in the context of the entire \textit{Kepler} stellar sample, is displayed in Fig. \ref{loggteff}. With such a working sample at hand, a careful treatment was applied to the LCs, using the so-called co-trending basis vectors provided by the \textit{Kepler} archive (see \citealt{2012Smith}; \citealt{2012Stumpe}; \citealt{2010Twicken}), to remove  systematic long-term trends originating from instruments, the environment, the detector, or effects caused by the re-orientation of the spacecraft after each $\sim$ 90 days. To remove outliers and  prepare the LCs for the analysis using spot modelling, we applied the method developed by \citet{2013Medeiros}, a procedure that is able to identify discontinuities in the LCs, similar to that used by \citet{2011Basri}. From this point on, an LC was considered to be fully treated, and its spot modelling analysis could be performed.

\section{The autocorrelation method}\label{sec:autocorr}

In this work, we follow the same procedure developed by \citealt{2014Lanza} to estimate surface DR. First, we applied an autocorrelation function (ACF) to check the stability of the photometric signal. Indeed, an important feature of the ACF is that it exhibits an oscillatory behavior with regularly spaced peaks; then, the coherence of a photometric signal can be estimated by the relative height of successive peaks in the ACF \citep{2014Lanza}. A crucial step in our analysis was the search for photometric signal stability for all 881 LCs constituting our initial working sample. From such analyses, we identified 17 stars presenting unambiguous stable photometric signals, indicating  rotational modulation. Nevertheless, in spite of the fact that a significant DR can be detected when the relative height of the second maximum in the ACF is at least 0.6-0.7 \citep{2014Lanza}, we have considered a few stars whose ACF has a peak ratio small than this threshold because the Monte Carlo Markov chain (MCMC) analysis points for a significant DR for them. These stars with less good ACFs are flagged by a dagger in Table \ref{tab1}. Indeed,  the ACF has been widely used in the study of photometric signals due to its ability to provide a good estimate of the average period variability, including stellar rotation period (e.g., \citealt{2013McQuillan}; \citealt{2012Affer}). Then, for these 17 stars with sufficiently stable signals, we applied spot modelling (\citealt{2014Lanza}) to seek individual spot rotation periods. The method of spot modelling is based on two spots and was applied with a Bayesian information criterion (hereafter BIC) to initially choose intervals of the time series presenting evidence of differential rotation with starspots of almost constant areas. The initial and final times $t_{1}$ and $t_{2}$, respectively, of those intervals are given in Table \ref{tab2}, together with the BIC computed values for each of the 17 stars. Indeed, $t_{1}$ and $t_{2}$ are defined in Barycentric Kepler Julian Day (BKJD). Even if the time intervals are particularly small, the spot modelling is able to give us a valid signal of DR, as many other authors (e.g., \citealt{2006ApJ...648..607C}; \citealt{2007AN....328.1037F}) have proven in previous studies. Readers are referred to \citet{2014Lanza} for a complete discussion of the ACF and the spot modelling procedure. Nevertheless, let us underline an important aspect, previously considered by different authors (e.g., \citealt{2015Davenport}; \citealt{2009Jeffers}), in the context of the present procedure. In the applied two-spot modelling, we cannot constrain the total number of starspots on the stellar surface, which, as noted by \citet{2015Davenport}, may reflect two groups of spots or even many small spots across the entire stellar surface.
 
The LCs and the oscillatory behaviour of the ACF for these 17 stars are shown in Fig. \ref{FigLCAF1} of Appendix A. The blue vertical solid lines display the initial and final times $t_{1}$ and $t_{2}$ of the intervals considered for the MCMC analysis. We then applied the procedure by \citet{2014Lanza} to compute the spot rotation period $P$, the mean values of the individual spot rotation periods $P_{1}$ and $P_{2}$ and their respective colatitudes, $\theta_1$ and $\theta_2$, and the relative amplitude of the DR, $\Delta P/P$, where $P = (P_{1} + P_{2})/2$. The a posteriori distributions of the rotation periods $P_{1}$ and $P_{2}$ of the two spots for all 17 stars, as derived from MCMC, are given in Fig. \ref{map}. The standard deviations of $\Delta P/P$ were also estimated by a model that assumes that starspots are not evolving along the fitted interval. Starspot evolution can limit our accuracy in measuring differential rotation at $\Delta \Omega \sim	1/t_{\rm evol}$, where $t_{\rm evol}$ is the evolutionary timescale, or even mimic a differential rotation signal in the worst cases (see \citealt{2015Aigrain}).

\begin{table}
\centering
\caption{\label{tab2}Initial and final times of the intervals considered for the MCMC analysis, together with the BIC computed values.}
\begin{tabular}{lccc}
\hline  \hline 
\multicolumn{1}{c}{Star} & $t_{1}$ & $t_{2}$ & BIC  \\
\multicolumn{1}{c}{(KIC \#)}		&  (BKJD)   & (BKJD)  &         \\
 \hline \hline 
2831979 	& 	1212.163 	& 	1255.382 	& 	4.097 	\\
4820062 	& 	735.384 	& 	761.804 	& 	18.778 	\\
5781991 	    & 	131.513 	& 	164.984 	& 	5.300 	\\
5956717 	& 	1182.758	& 1207.585 	& 	8.027 	\\
6143158 	& 	411.224	& 	439.158 	& 	17.008 	\\
6836955 	& 	261.205	& 	293.591 	& 	10.65 	\\
7430659 	& 	863.035 	& 899.592 	& 	3.261 	\\
8024188 	& 	1201.311 	& 	1240.893 	& 	7.467 	\\
8037792 	& 	634.978 	& 676.926 	& 	16.325 	\\
8495770 	& 	844.746 	& 880.383 	& 	19.635 	\\
9996105 	& 	264.290 	& 	336.132 	& 	3.183 	\\
10079452 	& 	448.517 	& 	499.623 	& 	8.413 	\\
10279927 	& 	1216.781 	& 1239.504 	& 	18.279 	\\
10460082 	& 	416.803 	& 	442.203 	& 	11.926 	\\
10514649 	& 	205.096 	& 	247.352 	& 	4.528 	\\
11199277 	& 	1419.912 	& 	1465.927 	& 	14.378 	\\
12520213 	& 	820.143 	& 	869.328 	& 	7.102 	\\		
\hline
	\end{tabular}
\end{table}	

\renewcommand{\thefootnote}{\fnsymbol{footnote}}
\begin{table*}
	\caption{\label{tab1}The stellar parameters and the results of MCMC analysis for our sample of 17 stars with traces of DR.}
	\begin{tabular}{lcccccccccc}
		\hline  \hline 
		\multicolumn{1}{c}{Star} & $P_{rot}$ & $\log g$ & $T_{\rm eff}$ & $P_{1}$ & $\sigma P_{1}$ & $P_{2}$ & $\sigma P_{2}$ & $\Delta P/P$ & $\sigma \Delta P/P$ & $\Delta \Omega$ \\
\multicolumn{1}{c}{(KIC \#)}		&  (d)     & (cm s$^{-2}$) &  (K)  &  (d)  &    (d)  &  (d)   &  (d)   &       &       & (d)  \\ \hline \hline 
		2831979\footnotemark[1] 	&	24.383	&	4.363	&	5783	&	24.433	&	9.597 x 10$^{-4}$&	22.660&	1.629 x 10$^{-3}$	&	0.0753	&	8.332 x 10$^{-5}$&0.02012 \\
		4820062\footnotemark[1] 	&	23.098	&	4.104	&	5699	&	22.634	&	1.296 x 10$^{-3}$&	24.771&	2.676 x 10$^{-3}$ 	&	0.0886	&	1.347 x 10$^{-4}$&0.02395\\
		5781991\footnotemark[2]	&	31.464	&	4.466	&	5796	&	28.947	&	3.150 x 10$^{-2}$&	33.006&	4.909 x 10$^{-2}$   &	0.1310	&	2.007 x 10$^{-3}$&0.02669\\
		5956717\footnotemark[1] 	&	23.297	&	4.213	&	5657	&	24.069	&	9.727 x 10$^{-4}$&	21.828&	6.862 x 10$^{-4}$	&	0.0976	&	5.440 x 10$^{-5}$&0.02679\\
		6143158\footnotemark[2]	&	23.155	&	4.509	&	5696	&	25.513	&	6.648 x 10$^{-3}$&	21.448&	2.218 x 10$^{-3}$	&	0.1731	&	3.243 x 10$^{-4}$&0.04667\\
		6836955\footnotemark[2]	&	26.354	&	4.61	&	5590	&	24.000	&	9.132 x 10$^{-3}$&	26.724&	5.223 x 10$^{-3}$	&	0.1074	&	4.371 x 10$^{-4}$&0.02668\\
		7430659\footnotemark[1] 	&	28.388	&	4.268	&	5823	&	29.897	&	1.377 x 10$^{-3}$&	26.440&	2.351 x 10$^{-3}$   &	0.1227	&	1.027 x 10$^{-4}$&0.02748\\
		8024188\footnotemark[1] 	&	23.849	&	4.384	&	5829	&	23.528	&	1.587 x 10$^{-3}$&	22.754&	1.653 x 10$^{-3}$	&	0.0334	&	1.007 x 10$^{-4}$&0.00908\\
		8037792\footnotemark[1]\footnotemark[2] &	23.400	&	4.285	&	5666	&	24.797	&	3.245 x 10$^{-3}$&	23.132&	2.280 x 10$^{-3}$	&	0.0695	&	1.712 x 10$^{-4}$&0.01823\\
		8495770\footnotemark[1]\footnotemark[2] 	&	25.634	&	4.545	&	5688	&	26.244	&	2.802 x 10$^{-3}$&	25.832&	1.605 x 10$^{-3}$	&	0.0158	&	1.250 x 10$^{-4}$&0.00383\\
		9996105	&	28.683	&	3.990	&	5815	&	30.006	&	1.936 x 10$^{-2}$&	26.012&	1.881 x 10$^{-2}$   &	0.1426	&	1.032 x 10$^{-3}$&0.03215\\
		10079452\footnotemark[1] &	26.261	&	4.030	&	5812	&	25.092	&	7.177 x 10$^{-3}$&	25.842&	3.767 x 10$^{-3}$   &	0.0294	&	3.230 x 10$^{-4}$&0.00726\\
		10279927 &	24.018	&	4.508	&	5638	&	23.429	&	7.529 x 10$^{-4}$&	22.373&	8.778 x 10$^{-4}$	&	0.0461	&	5.166 x 10$^{-5}$&0.01265\\
		10460082\footnotemark[1] &	26.027	&	4.554	&	5835	&	24.321	&	2.090 x 10$^{-3}$&	28.001&	3.333 x 10$^{-3}$	&	0.1407	&	1.610 x 10$^{-4}$&0.03396\\
		10514649\footnotemark[1] &	24.205	&	4.388	&	5651	&	24.875	&	1.740 x 10$^{-2}$&	22.694&	1.084 x 10$^{-2}$   &	0.0917	&	9.016 x 10$^{-4}$&0.02428\\
		11199277\footnotemark[1] &	29.339	&	4.493	&	5638	&	30.356	&	3.538 x 10$^{-3}$&	28.019&	1.479 x 10$^{-3}$	&	0.0801	&	1.366 x 10$^{-4}$&0.01726\\
		12520213 &	25.318	&	4.457	&	5679	&	25.270	&	3.361 x 10$^{-3}$&	23.967&	2.941 x 10$^{-3}$	&	0.0529	&	1.862 x 10$^{-4}$&0.01352\\
		\hline
	\end{tabular}
\footnotemark[1]{Stars with manifestation of DR, which are in common with \citet{Reinhold}.}\\
\footnotemark[2]{Stars with ACFs lower than the threshold 0.6-0.7.}\\
\end{table*}

\section{Results}\label{sec:results}

\begin{figure}
	\centering 
	\includegraphics[width=0.48\textwidth]{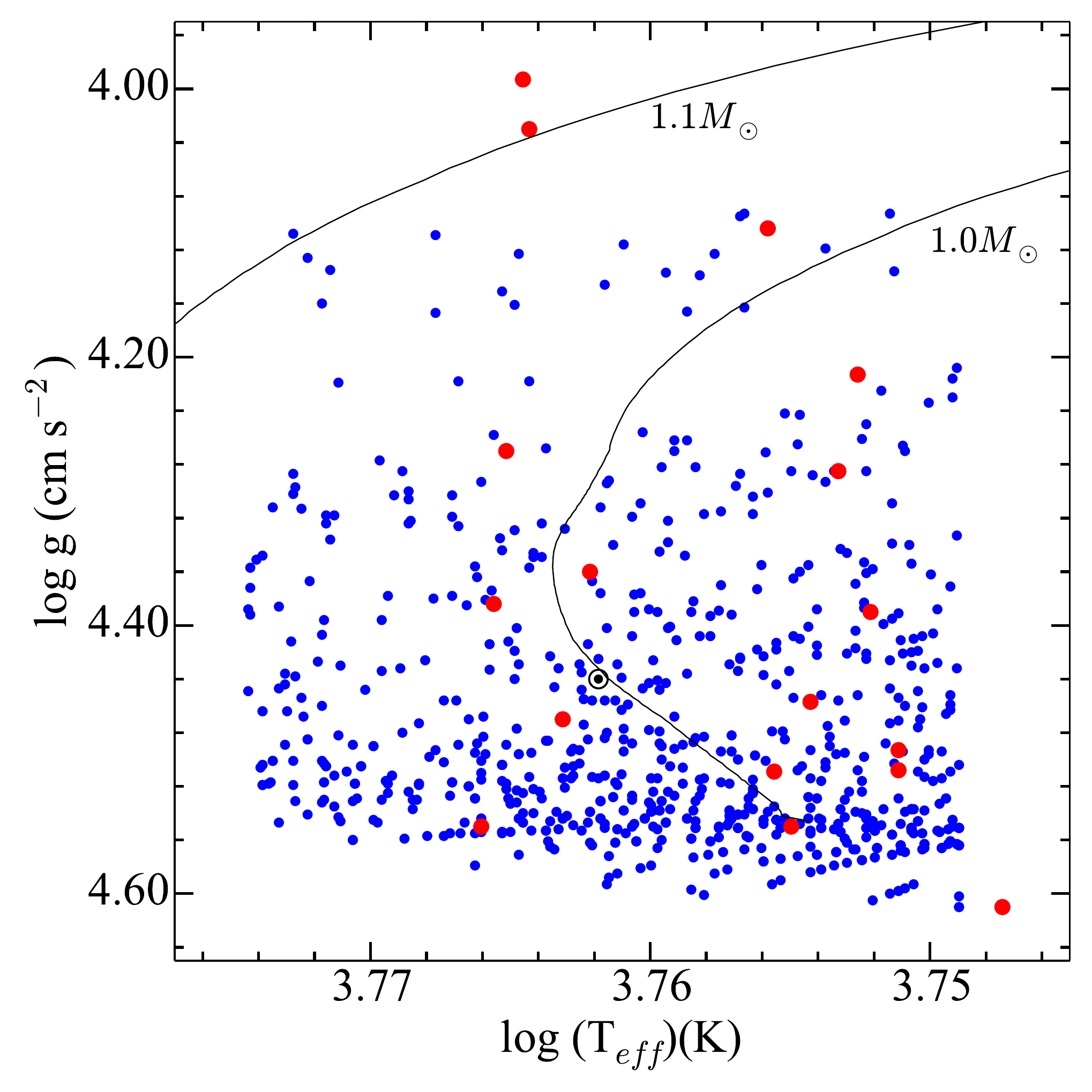} 
	\caption[]{Distribution of Sun-like stars with measured DR in the $\log g$ and $T_{\rm eff}$ diagram, represented by red circles. Blue circles indicate stars of the original working sample without traces of DR. The evolutionary tracks are from \citet{2012Silvia}.}
	\label{hrsel1}
\end{figure}

The main results of the present study are given in Table \ref{tab1}, which lists the mean values of the individual spot rotation periods $P_1$ and $P_2$, the relative amplitude of the DR lower limit, $\Delta P/P$ and the amplitude of the DR expressed as the frequency difference between the spots frequencies, $\Delta \Omega$. Table \ref{tab1} lists also the stellar parameters $P_{rot}$, $\log g$ and $T_{\rm eff}$. Fig. \ref{hrsel1} displays, in the $\log g$ vs. $T_{\rm eff}$ diagram, the locations of the  881 stars defined in our selection criteria, namely stars showing physical properties that are approximately equal to the Sun values, with 3.94 cm s$^{-2} < \log g <$ 4.94 cm s$^{-2}$, 5579 K  $< T_{\rm eff} < $ 5979 K, and the rotation period ranging into the solar values, from 23 days  $< P_{rot} <$ 33 days. In the referred figure, the red points represent the 17 stars having spot lifetimes long enough for the detection of DR patterns on the basis of our spot modelling method. Evolutionary tracks taken from \citet{2012Silvia} are overlayed to constrain the view of the mass range and evolutionary stage of the sample stars, with the position of the Sun indicated by the black symbol.

\begin{figure}
	\centering 
	\includegraphics[width=0.47\textwidth]{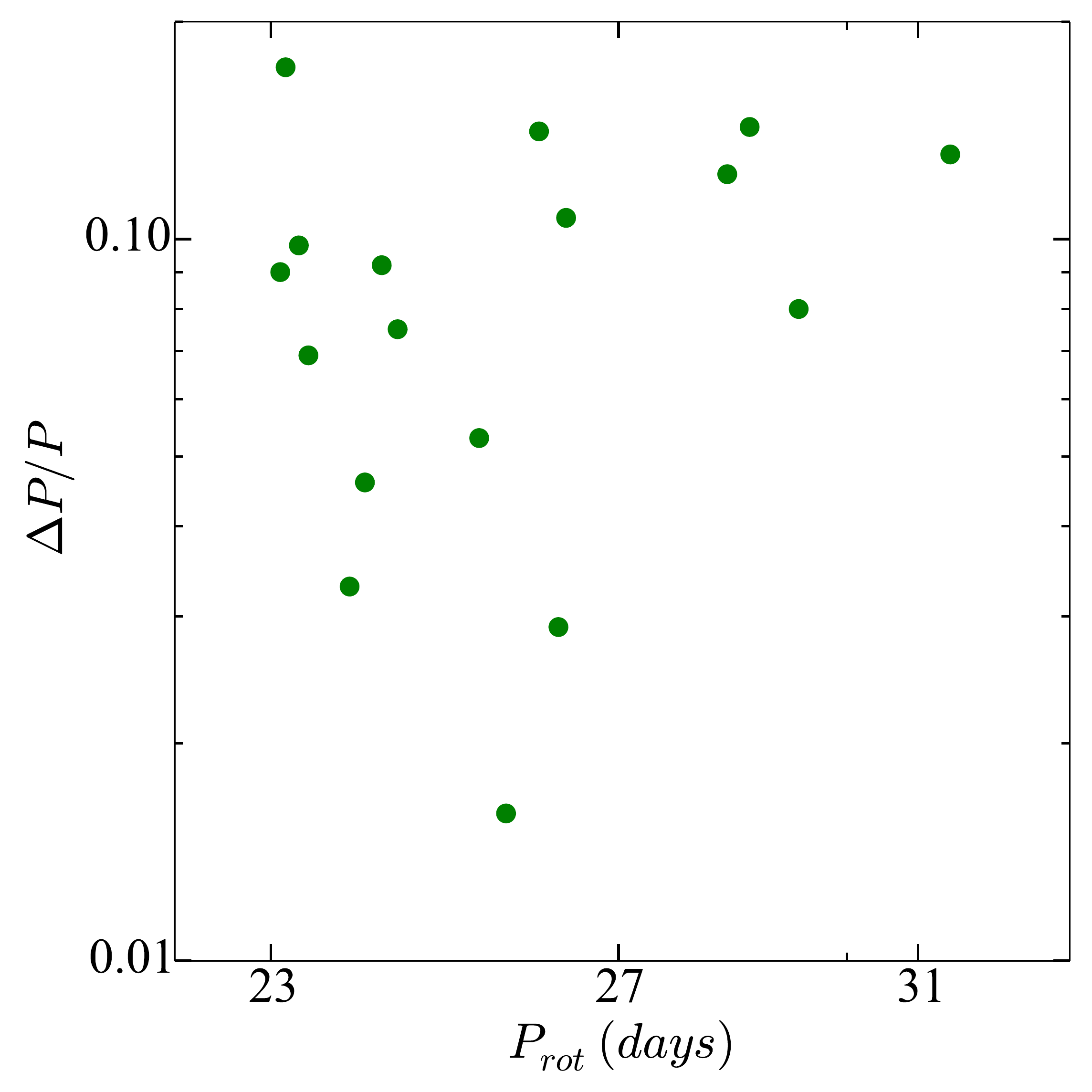} 
	\caption[]{The distribution of the relative amplitude $\Delta P/P$ versus the rotation period $P_{rot}$ for the 17 Sun-like stars with DR traces identified in the present study.}
	\label{hrsel2}
\end{figure}

We compared the present results with \citet{Reinhold}. Indeed, from our sample of 17 stars with measured amplitude of surface DR, 11 stars are found to be in common with those authors. For these common stars, \citet{Reinhold} detected the presence of multiple periods in their LCs, which were interpreted as the manifestation of DR, using a different approach based on the Lomb-Scargle periodogram. These stars are identified in Table \ref{tab1} with an asterisk. A simple comparison between the DR values of this small sample of targets in common provides no correlation between the values, although they are distributed in a similar range. As a more robust test, we applied Student’s t-test, which can be used to compare whether measures in one sample are paired with measures in another sample. According to this method, the null hypothesis assumes that the true mean difference between the two observations on each sample is zero; otherwise, the alternative hypothesis is considered. In this sense, the results of the paired t-test show that, because the -t$_{0.025}$(-2.228) $<$ t$_{computed}(-0.959)$ $<$ t$_{0.025}$(2.228) and because the p-value $>\,0.05$ (confidence level), we cannot reject the null hypothesis. Such a fact may reflect, in principle, the difference in the nature of the procedures applied in the search for DR traces. In addition, the compatibility between their ranges suggests that their information is valid at least up to an order of magnitude. 

Finally, we analysed the behaviour of the relative amplitude $\Delta P/P$ as a function of rotation period for our sample of 17 stars despite the narrow range of rotation periods considered in the present study, namely, from 23 to 33 days. Fig. \ref{hrsel2} displays the behaviour of $P_{rot}$ vs. $\Delta P/P$, from which one observes a soft trend of increasing $\Delta P/P$ towards longer rotation periods, paralleling the scenario found by different studies. For instance, as shown by \citet{2013Reinhold}, the relative DR shear increases with longer rotation periods, in agreement with previous observations (\citealt{2005Barnes}) and theoretical approaches (\citealt{2011Kuker}).

\section{Conclusions}\label{sec:concl}

Based on a simple two-spot model together with a Bayesian information criterion, we measured a lower limit on the amplitude of surface DR for 17 \textit{Kepler} Sun-like stars. For these stars, using \textit{Kepler} high-precision and evenly sampled photometric time series, it was possible to compute the spot rotation period $P$, the mean values of the individual spot rotation periods $P_1$ and $P_2$ and the relative amplitude of the differential rotation, $\Delta P/P$, where $P=(P_1 + P_2)/2$. These stars present a soft trend of the estimated relative amplitude, $\Delta P/P$, increasing with increasing rotation periods, in agreement with the scenarios found in the literature, from several observational studies of DR, based on different measurement approaches. 

In summary, although the art of measurements of the surface rotation of stars has now been mastered, with a high level of precision and maturity, the detection and measurement of stellar differential rotation remains a tricky subject. In the present study, using a spot-modelling procedure, we were able to detect surface differential rotation patterns in 17 stars with physical properties, including rotation, similar to the Sun. The portrait emerging from the present study points to a significant perspective: among Sun-like stars with surface rotation similar to the solar values, surface differential rotation appears to be a common phenomenon.

\section*{Acknowledgements}
 The research activity of the Observational Astronomy Board of the Federal University of Rio Grande do Norte (UFRN) is supported by continuous grants from CNPq and FAPERN brazilian agencies. We also acknowledge financial support from INCT INEspaço/CNPq/MCT. MLC, JPB and ADC acknowledge CAPES/PNPD fellowships. ICL and CEFL acknowledge  CNPq/PDE fellowships. RSB and FPC acknowledge
graduate fellowships from CAPES. D. B de Freitas also acknowledges financial support by the Brazilian agency CNPq (Grant No. 306007/2015-0). This paper includes data collected by the \textit{Kepler} mission. Funding for the \textit{Kepler} mission is provided by the NASA Science Mission directorate. All \textit{Kepler} data presented in this paper were obtained from the Mikulski Archive for Space Telescopes (MAST). We would like to thank the anonymous referee for the helpful comments that lead us to a substantial improvement of this manuscript.


\bibliographystyle{mnras}
\bibliography{biblio} 


\appendix
\section{Figures}	

\begin{figure*}
	\centering
	\includegraphics[scale=0.295]{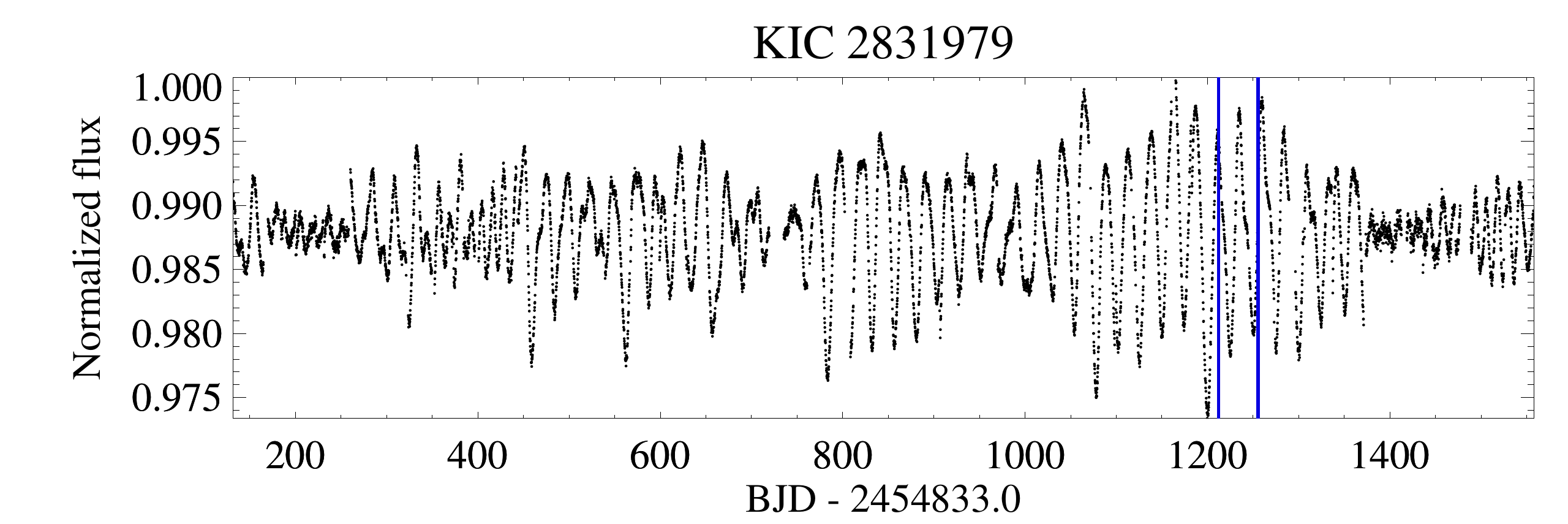}
	\includegraphics[scale=0.295]{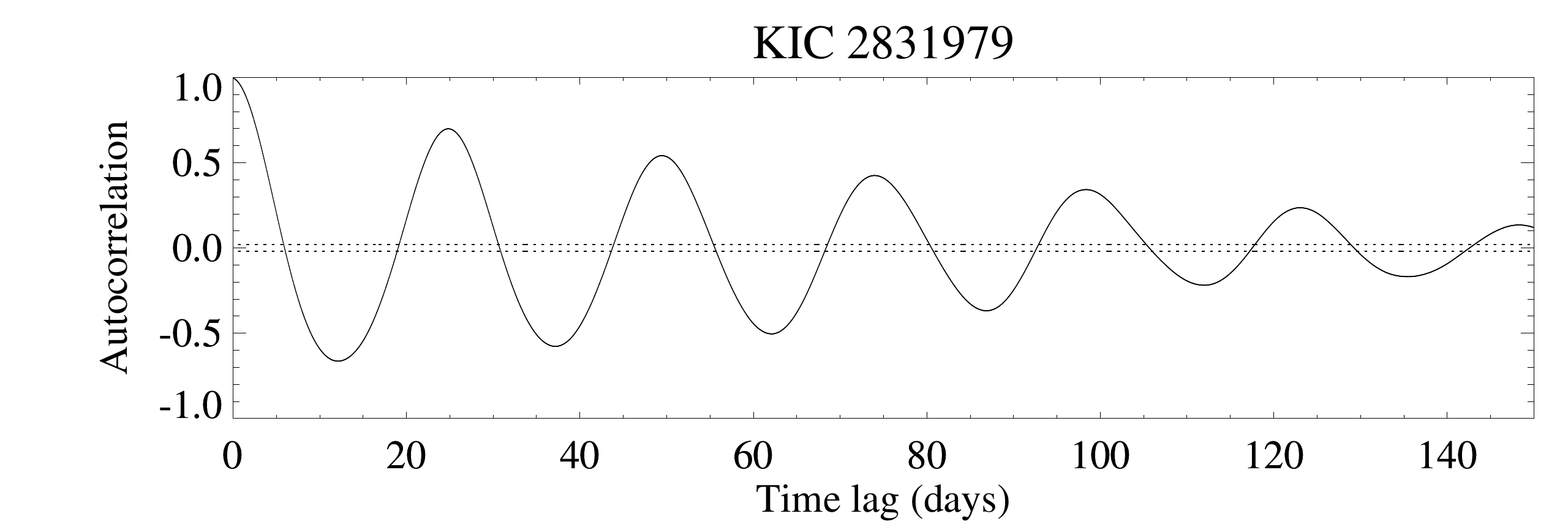}
	\includegraphics[scale=0.295]{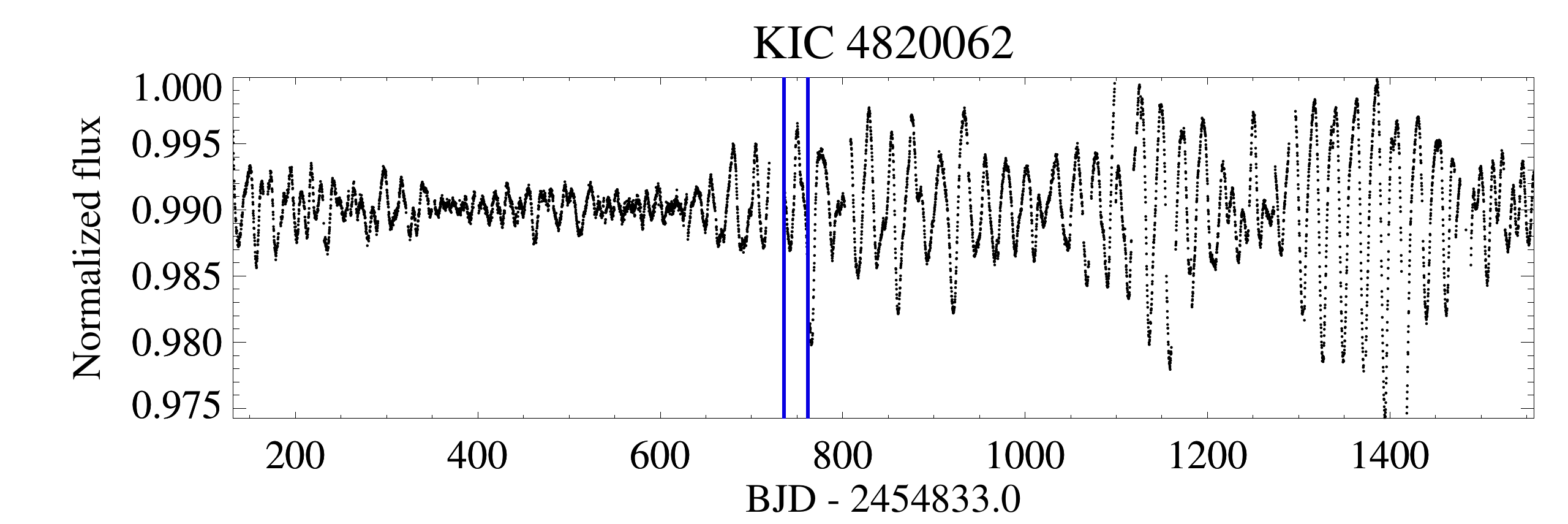}
	\includegraphics[scale=0.295]{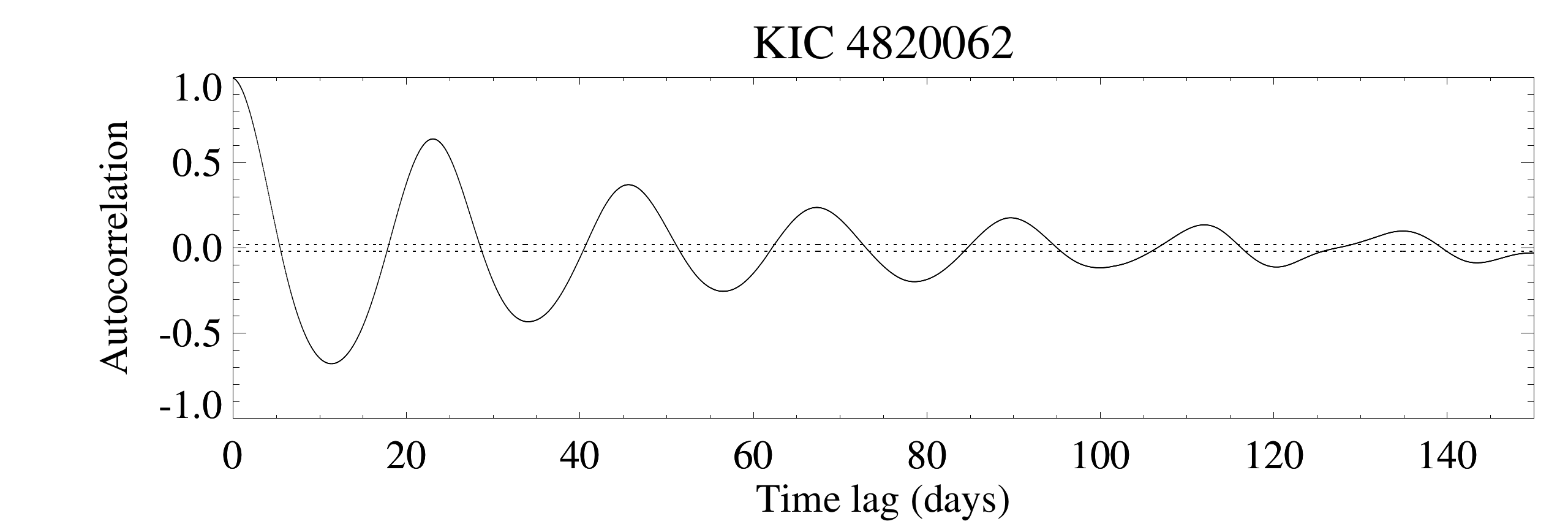}
	\includegraphics[scale=0.295]{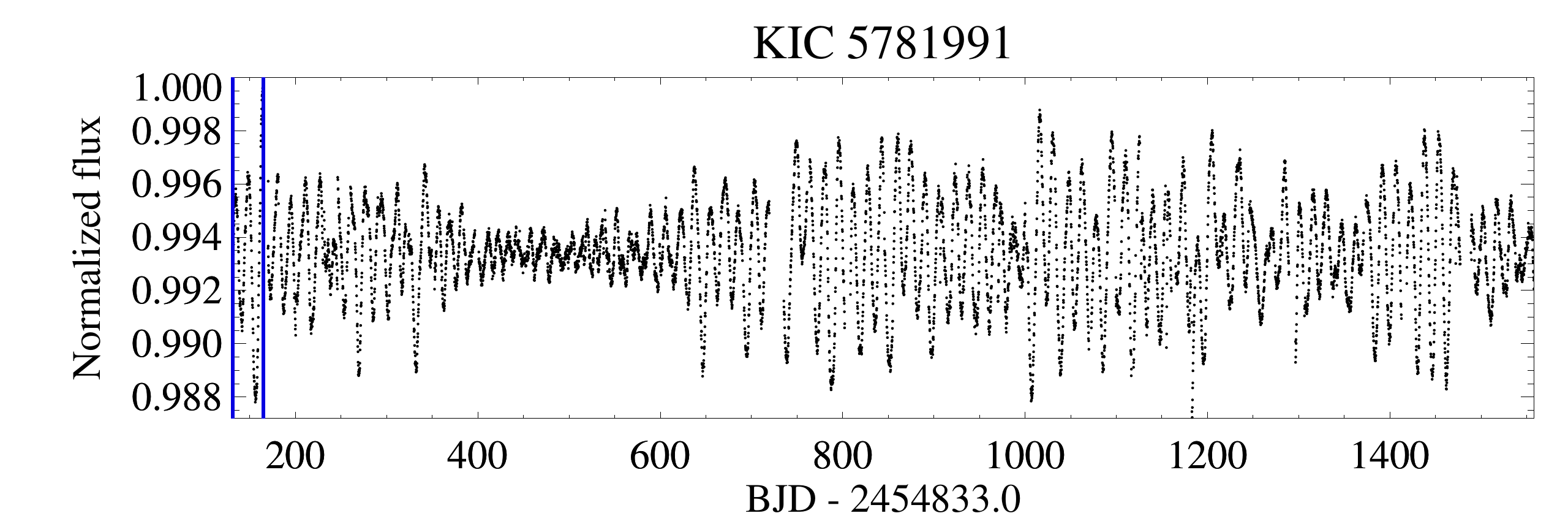}
	\includegraphics[scale=0.295]{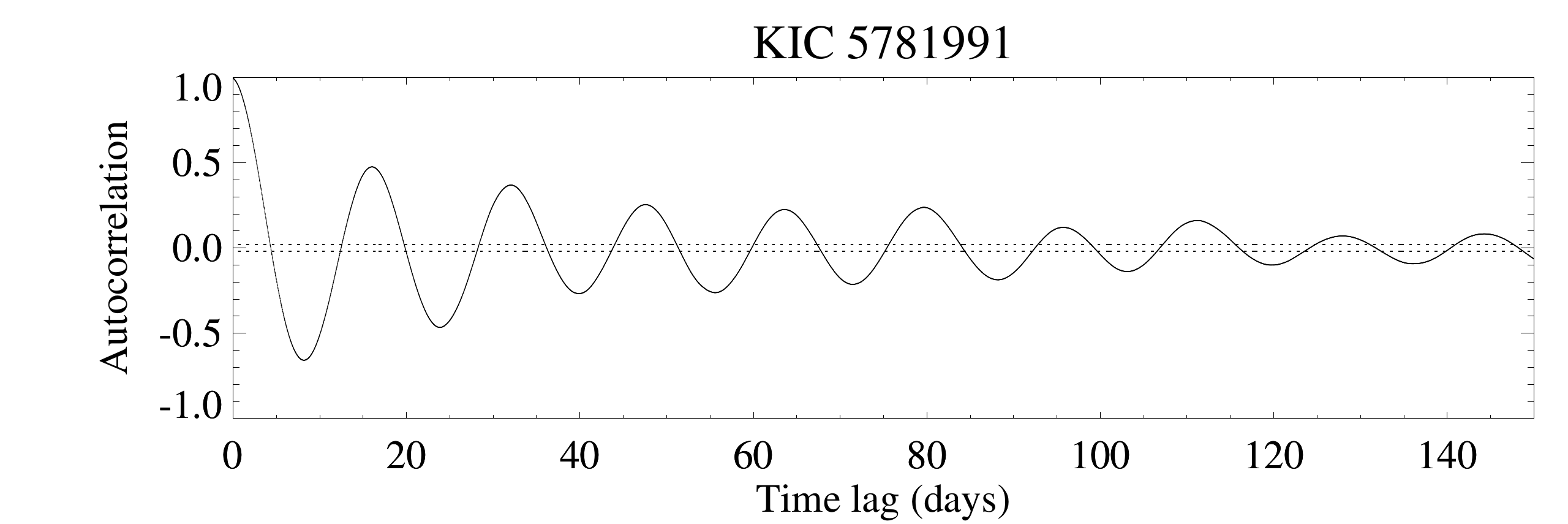}
	\includegraphics[scale=0.295]{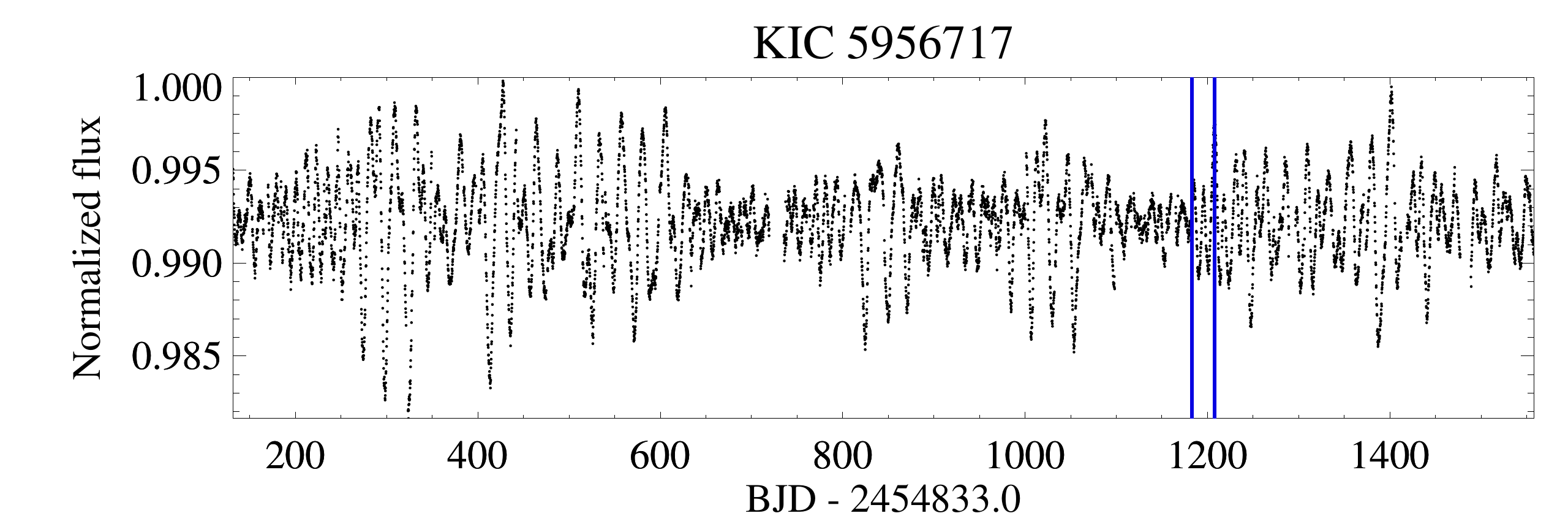}
	\includegraphics[scale=0.295]{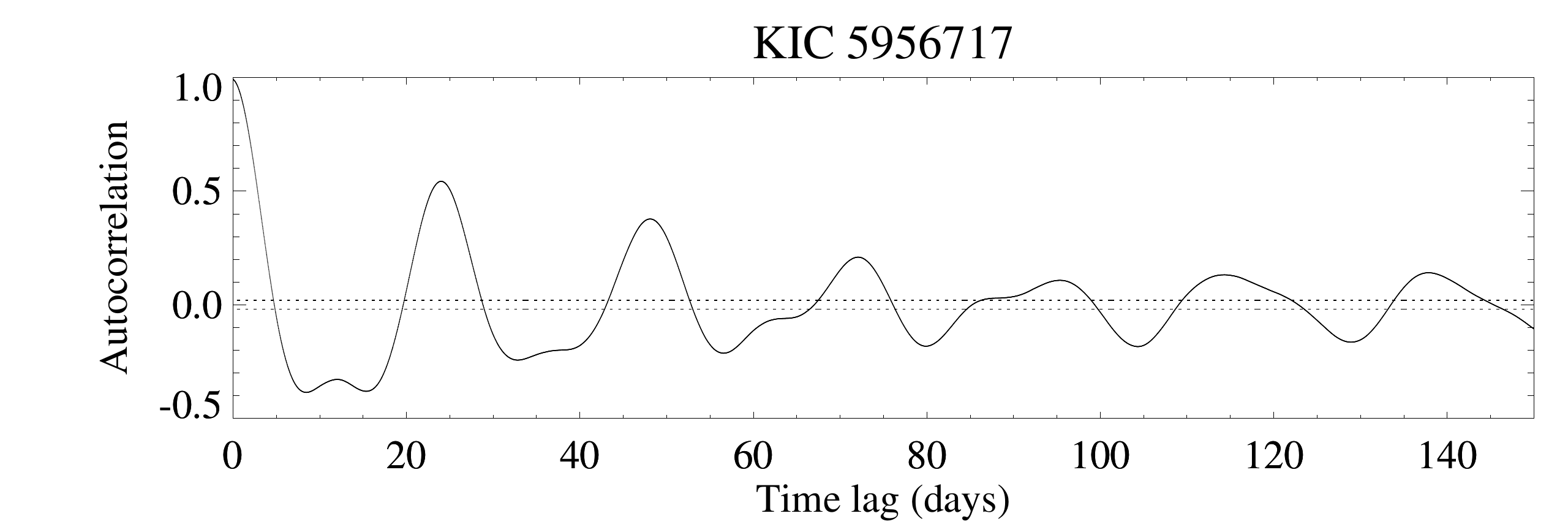}
	\includegraphics[scale=0.295]{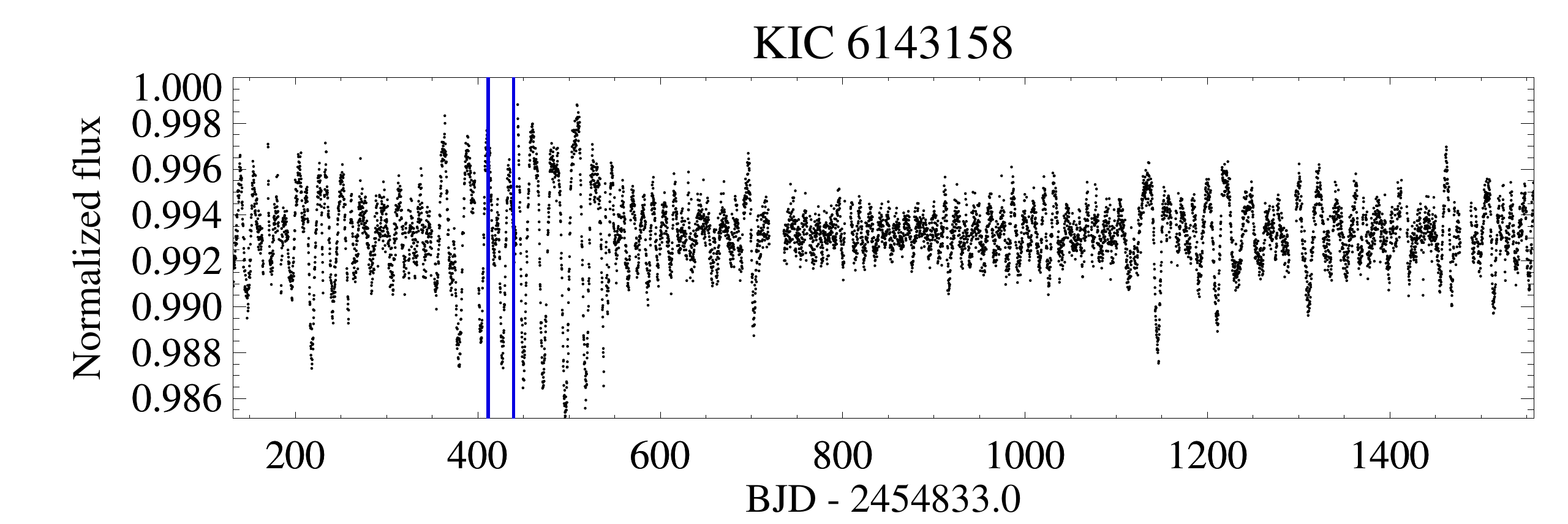}
	\includegraphics[scale=0.295]{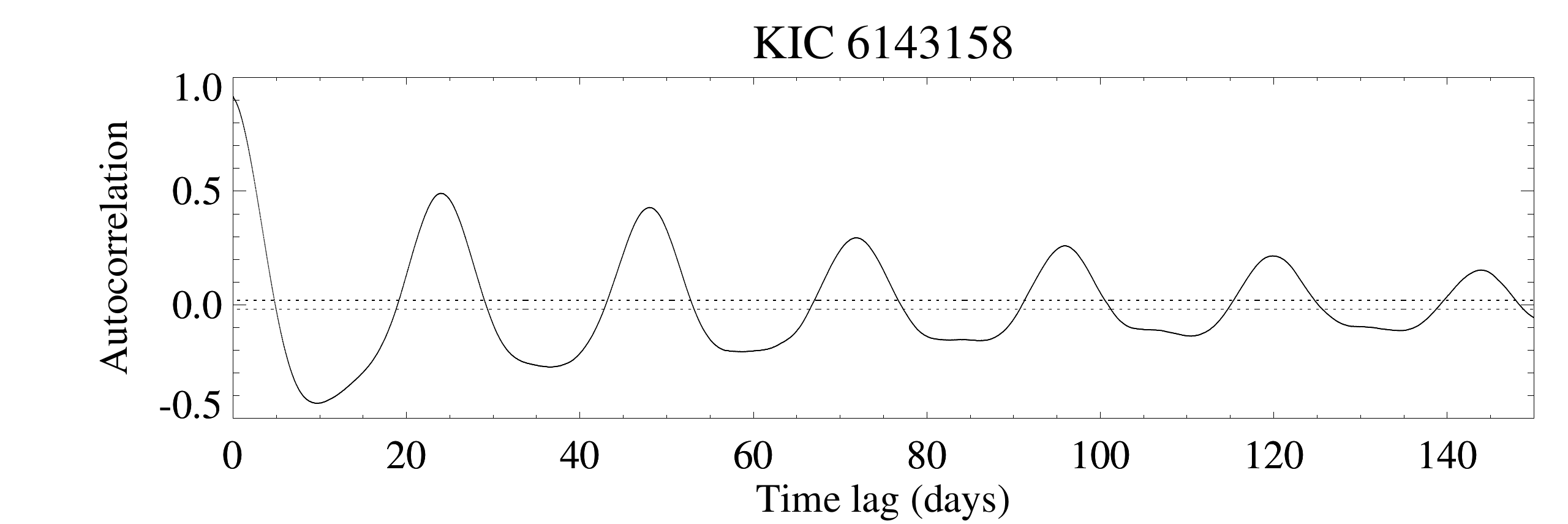}
	\includegraphics[scale=0.295]{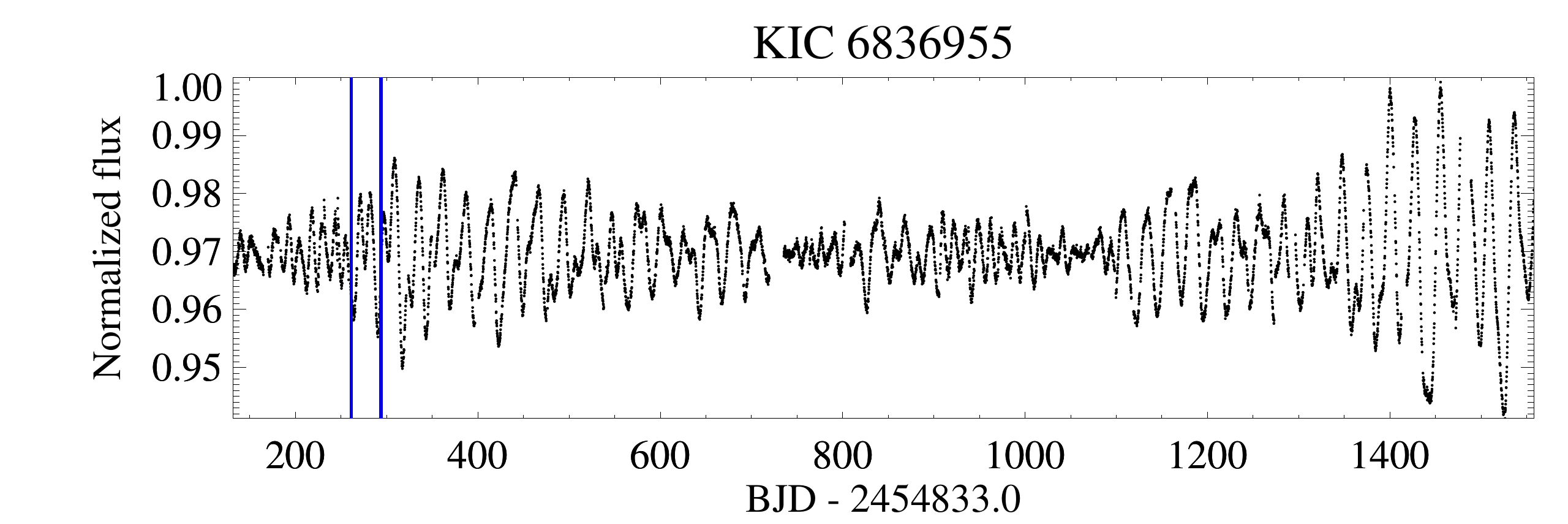}
	\includegraphics[scale=0.295]{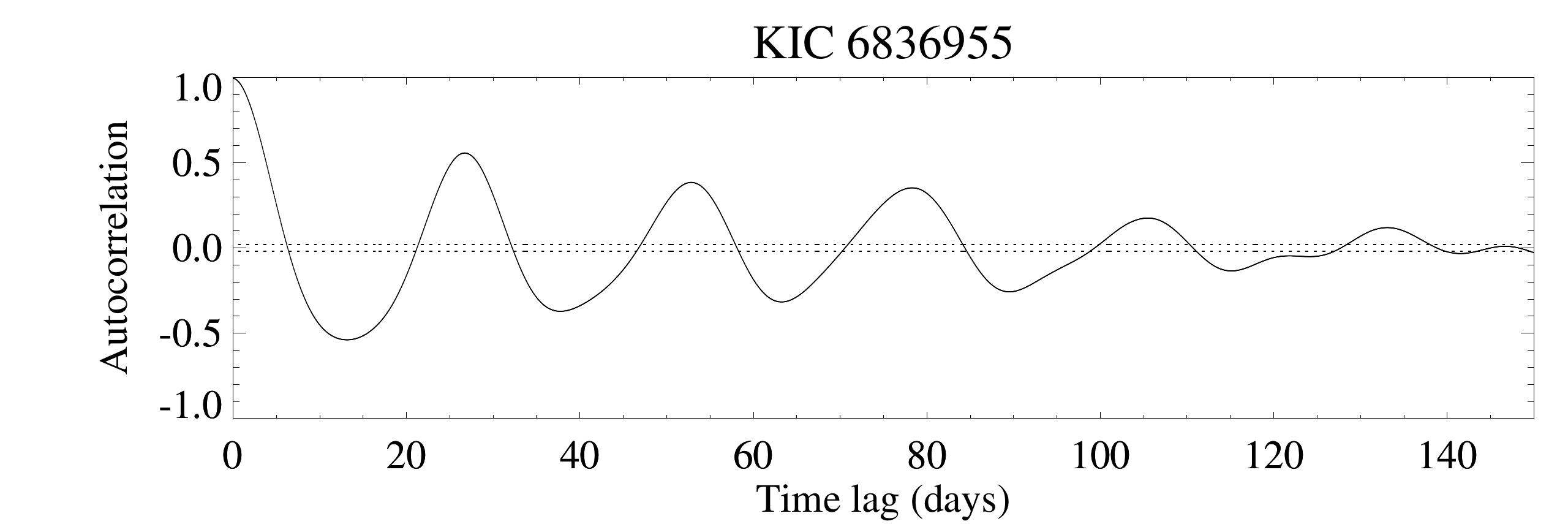}
	\includegraphics[scale=0.295]{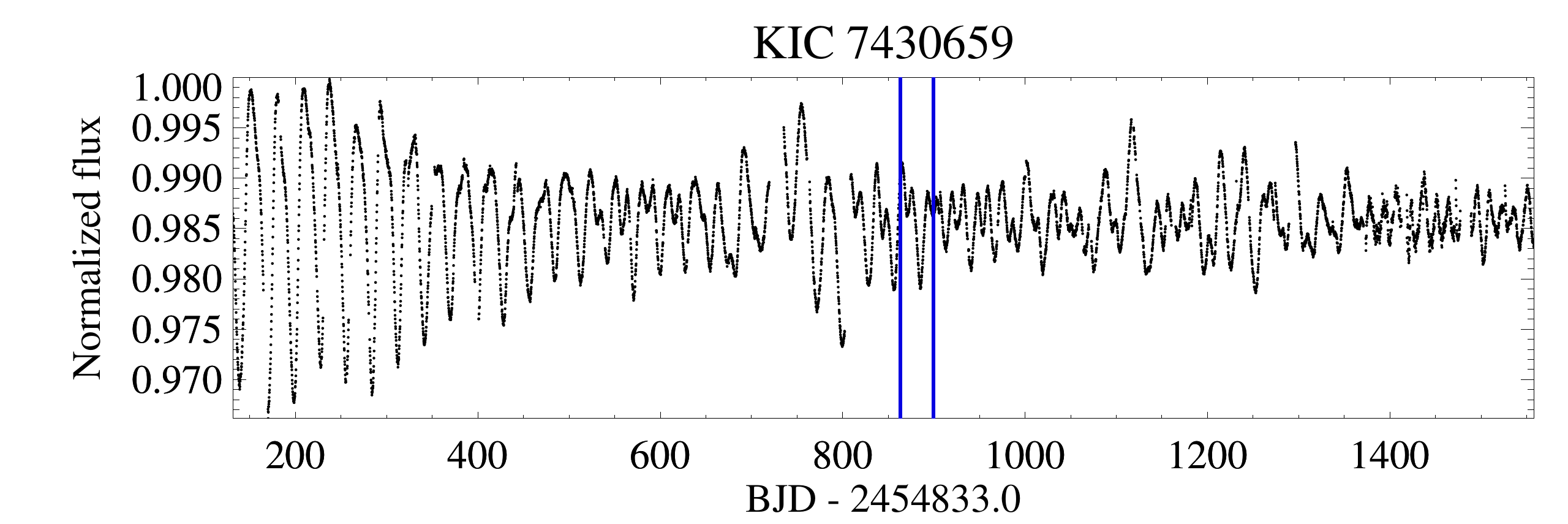}
	\includegraphics[scale=0.295]{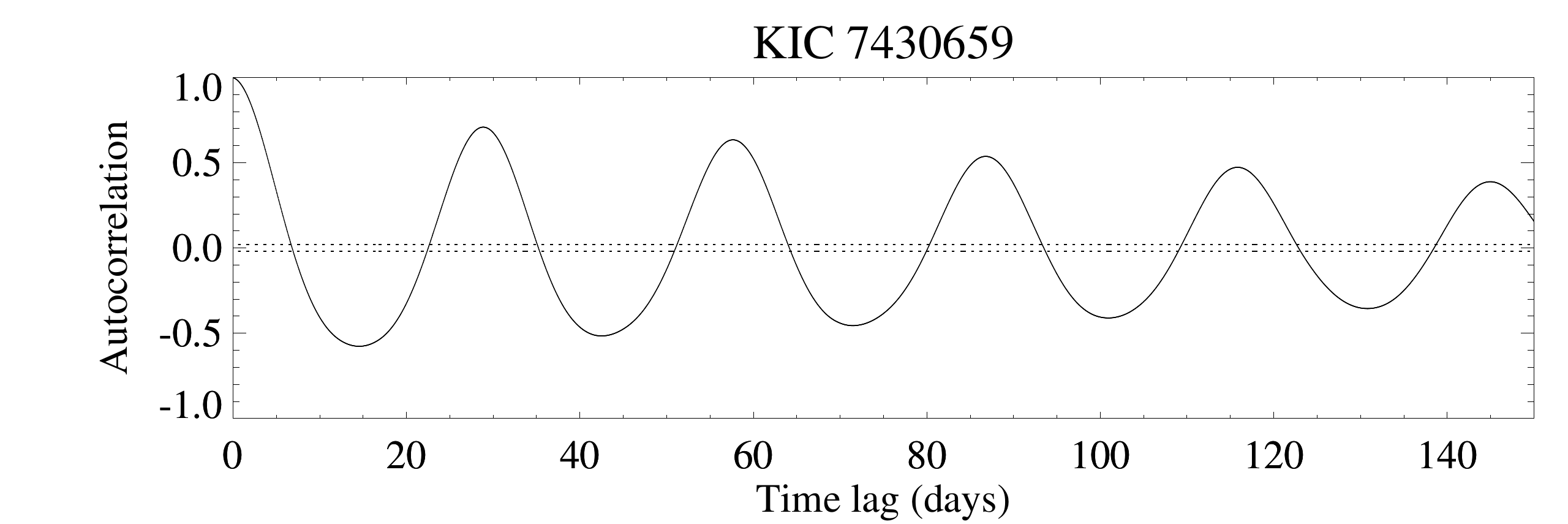}
	\includegraphics[scale=0.295]{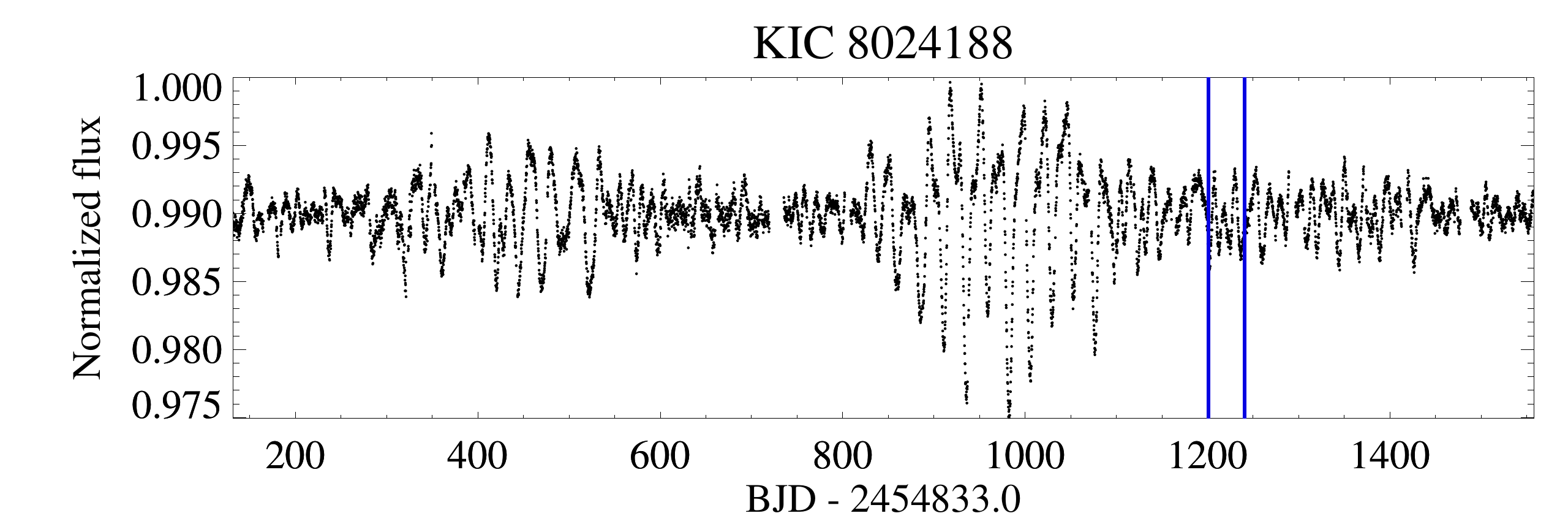}
	\includegraphics[scale=0.295]{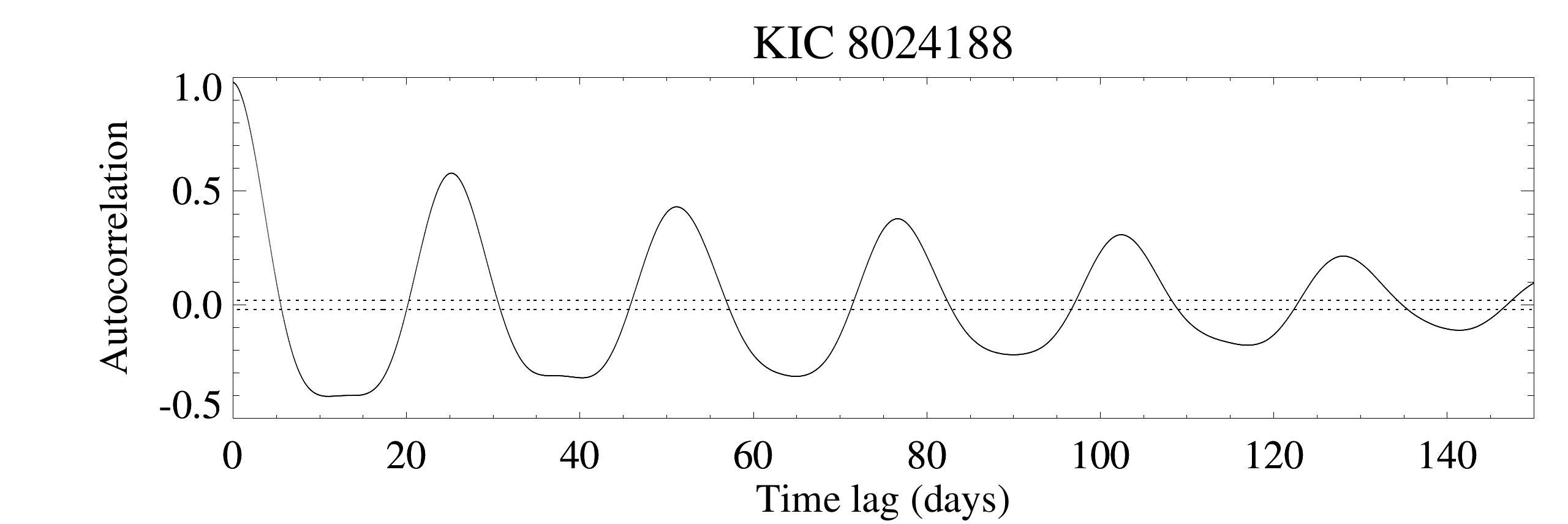}
	\caption{Left: Photometric times series of {\em{Kepler}} stars (from top to bottom) KIC 2831979, KIC 4820062, KIC 5781991, KIC 5956717, KIC 6143158, KIC 6836955, KIC 7430659, KIC 8024188, KIC 8037792, KIC 8495770, KIC 9996105, KIC 10079452, KIC 10279927, KIC 10460082, KIC 10514649, KIC 11199277 and KIC 12520213. The flux has been normalized to the maximum value observed along each time series. The vertical solid lines (in blue) display the initial and final times of the intervals considered for MCMC analysis (see Table \ref{tab2}).  Right: Autocorrelation functions of the LCs of the stars in our sample. The dotted lines indicate the interval corresponding to $\pm$$\sigma$, where $\sigma$ is one standard deviation of the autocorrelation as expected for a pure random noise with some degree of autocorrelation according to the large-lag approximation.}
	\label{FigLCAF1}
\end{figure*}

\begin{figure*}	
	\centering
	\includegraphics[scale=0.295]{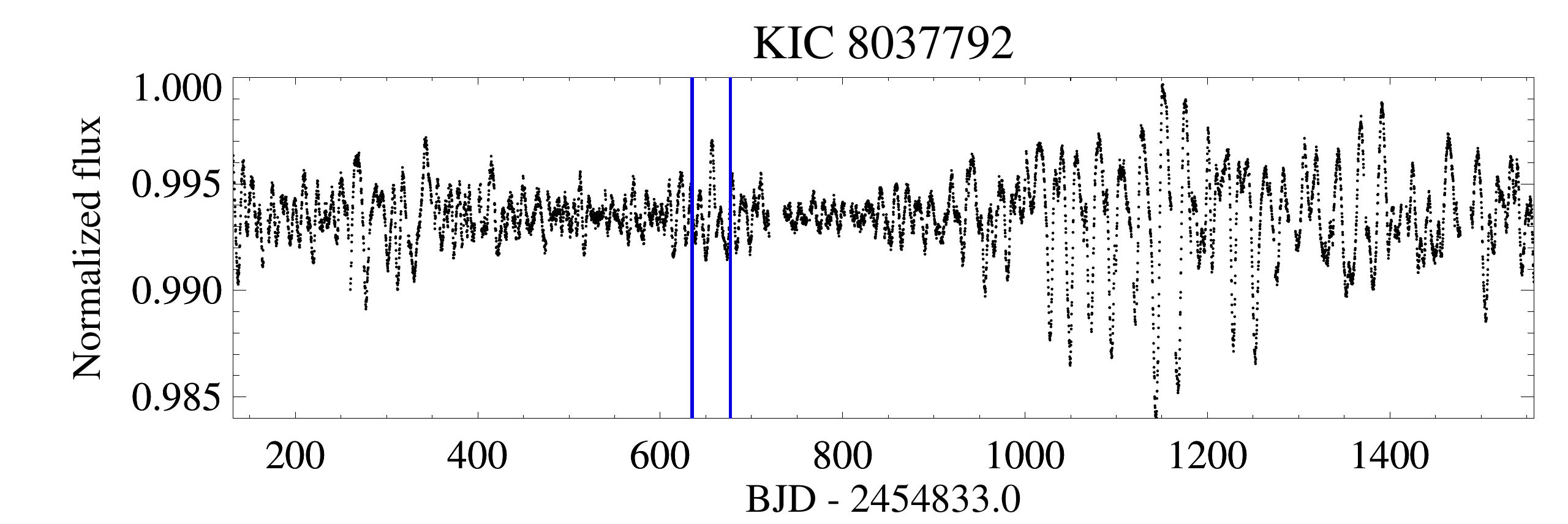}
	\includegraphics[scale=0.295]{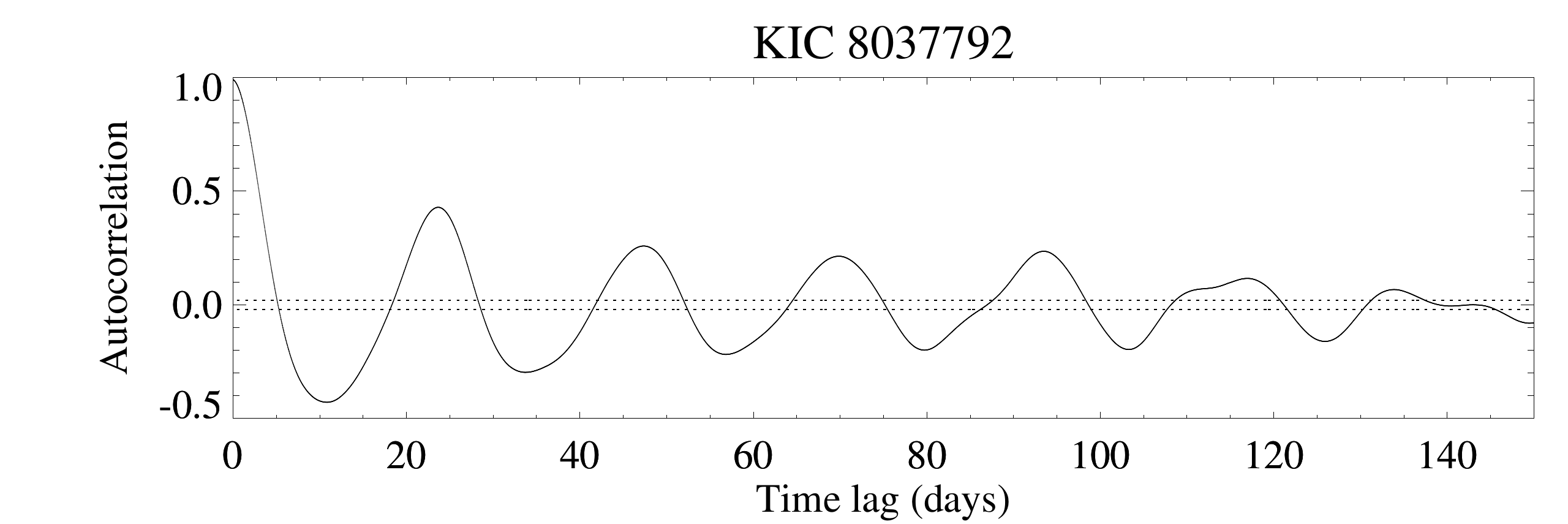}
	\includegraphics[scale=0.295]{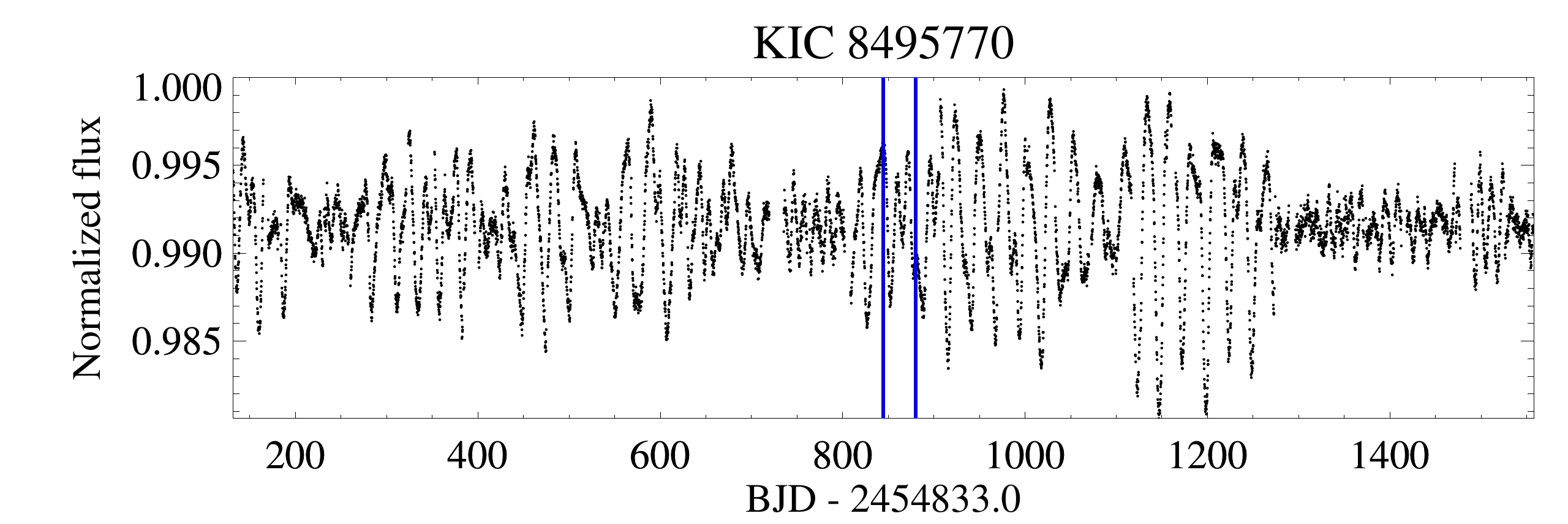}
	\includegraphics[scale=0.295]{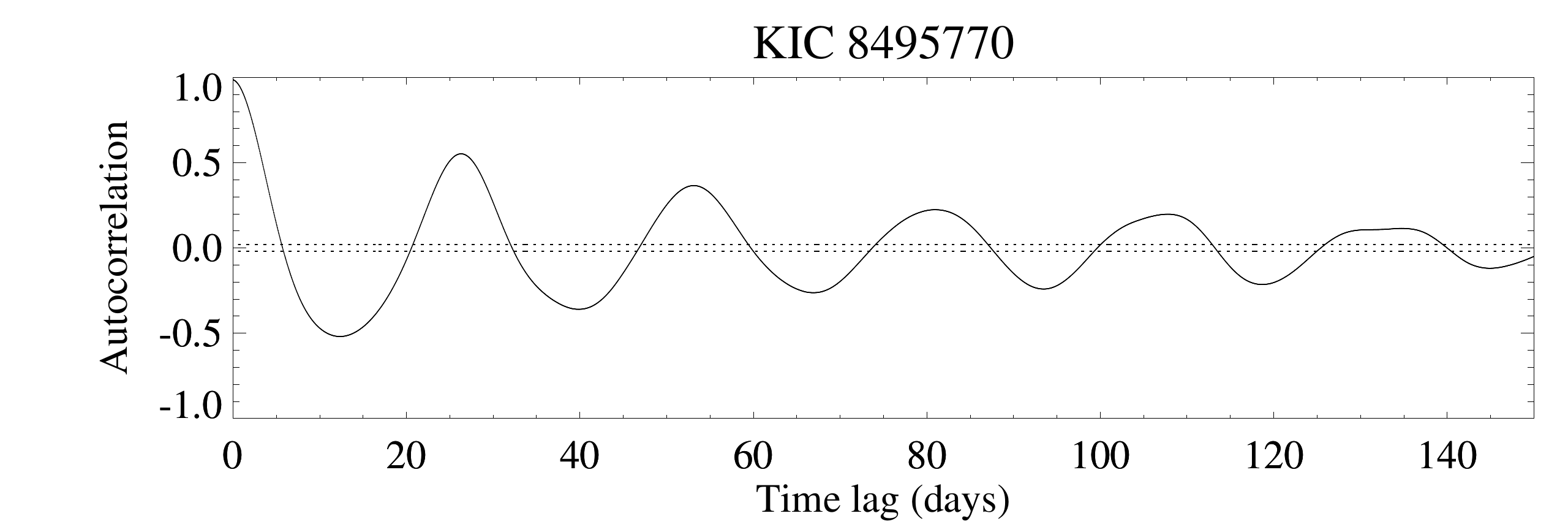}
	\includegraphics[scale=0.295]{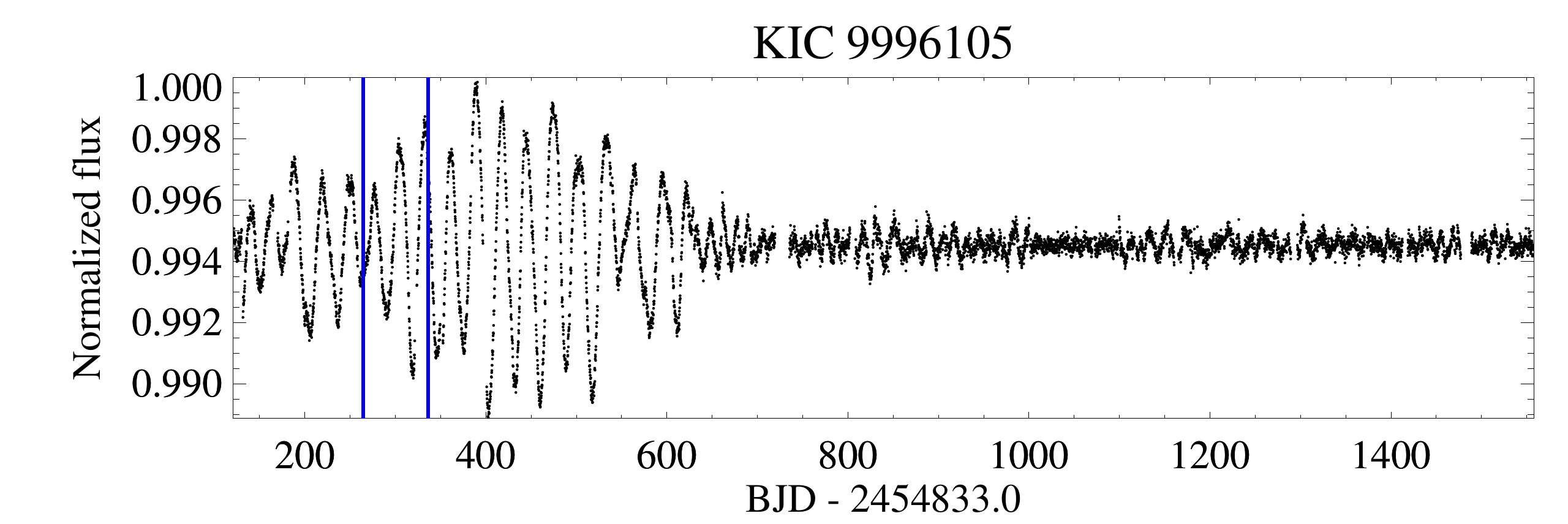}
	\includegraphics[scale=0.295]{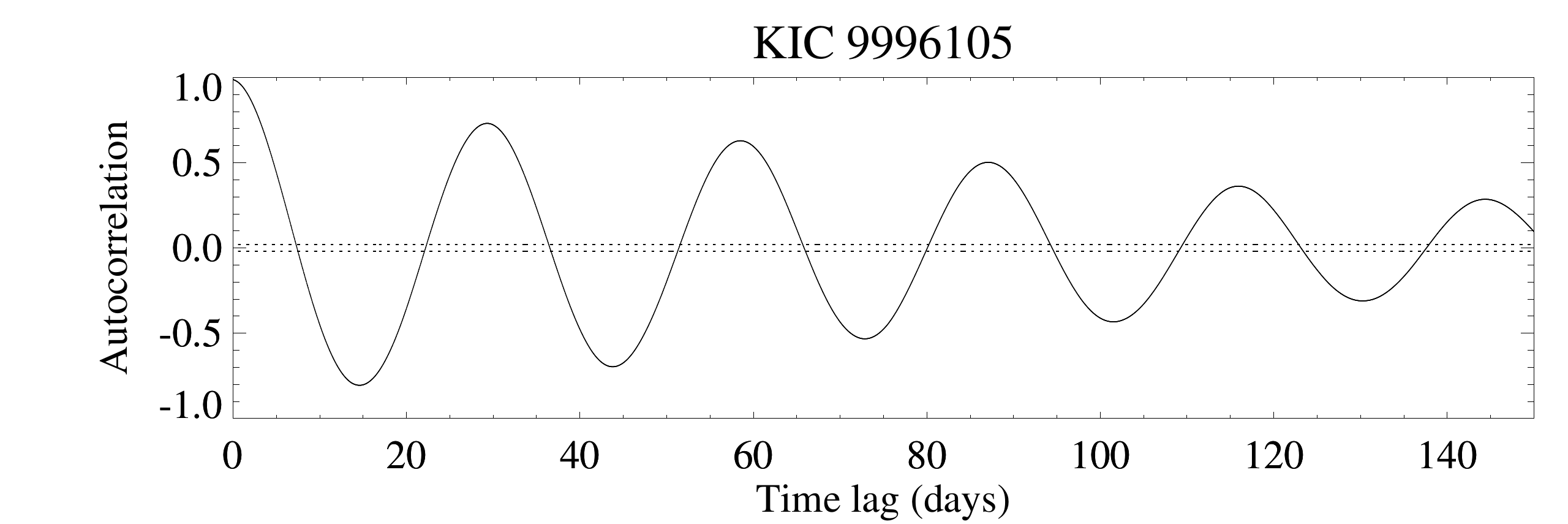}
	\includegraphics[scale=0.295]{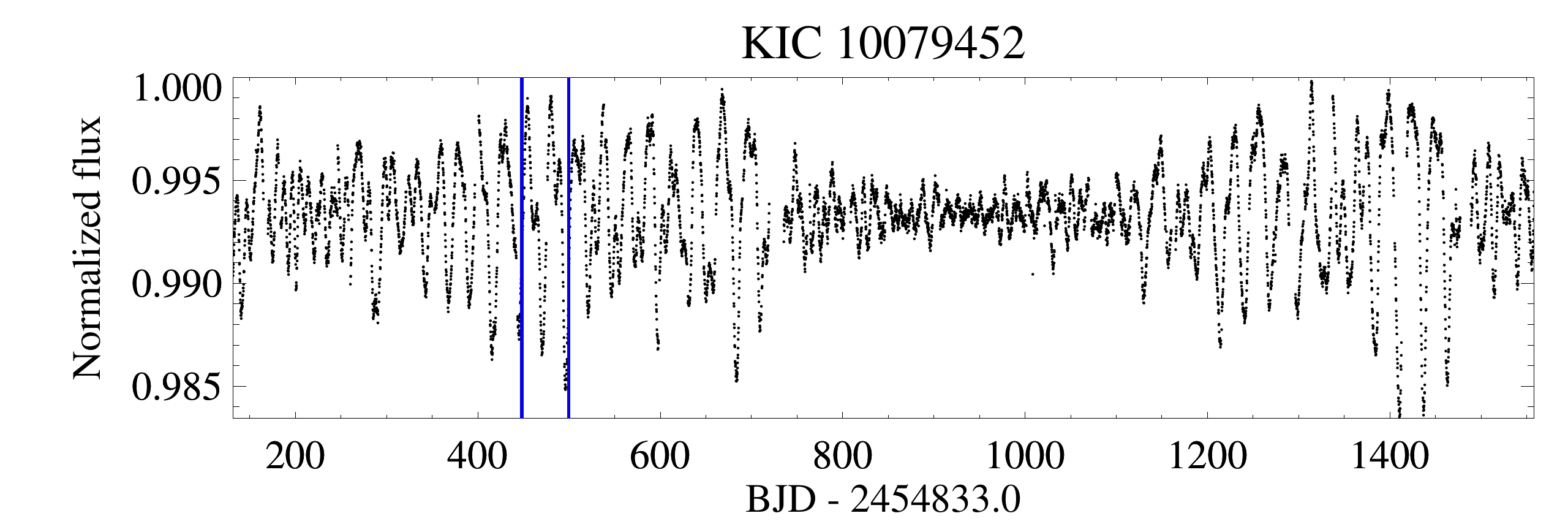}
	\includegraphics[scale=0.295]{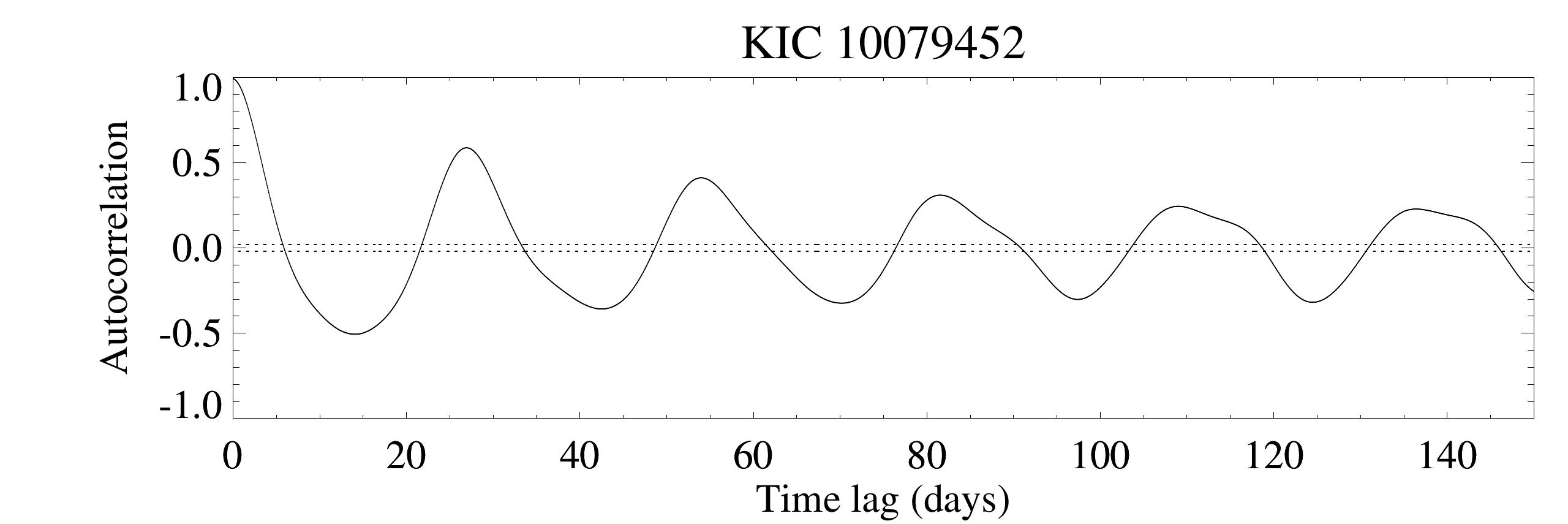}
	\includegraphics[scale=0.295]{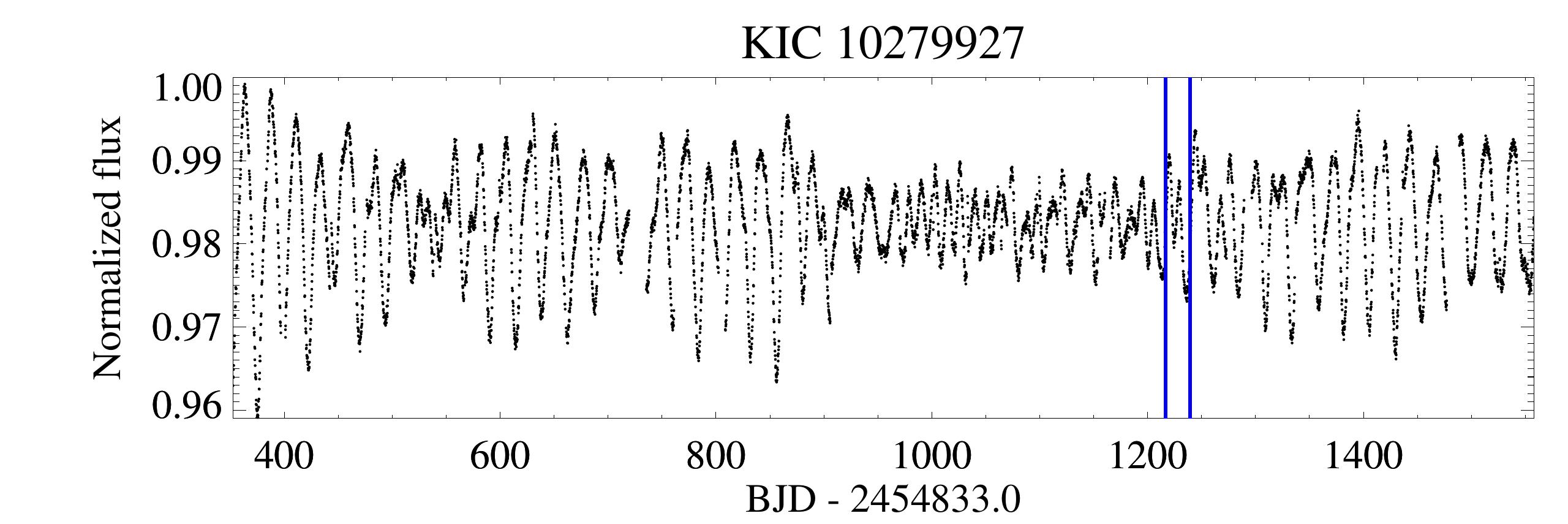}
	\includegraphics[scale=0.295]{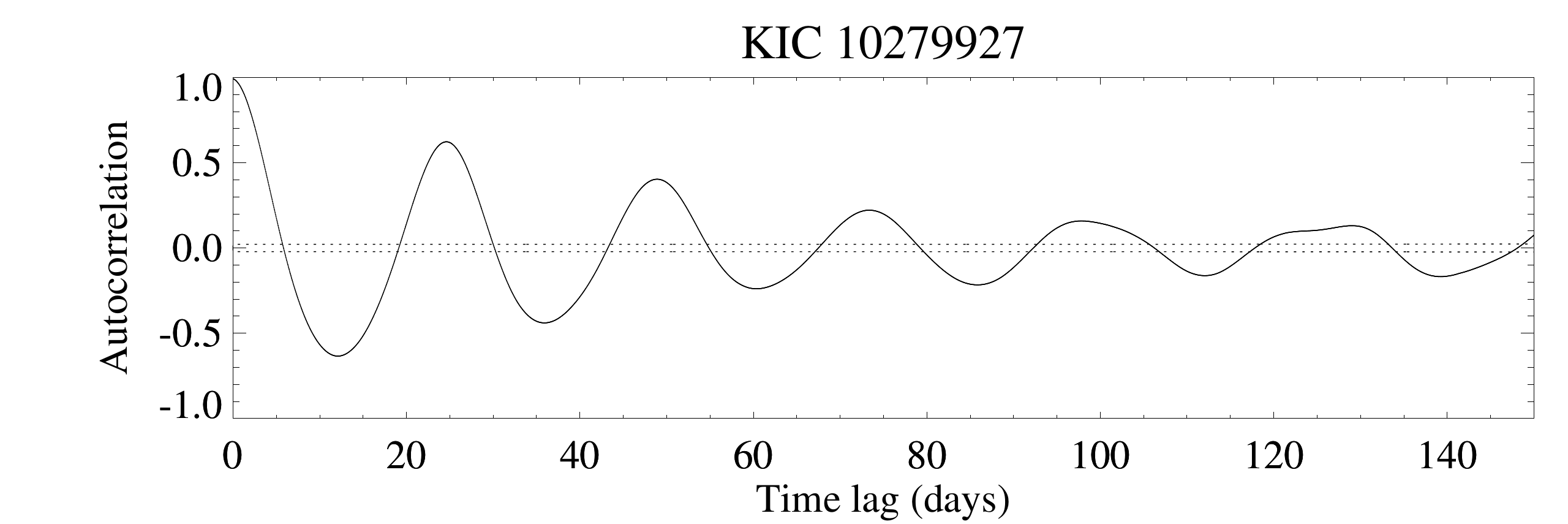}
	\includegraphics[scale=0.295]{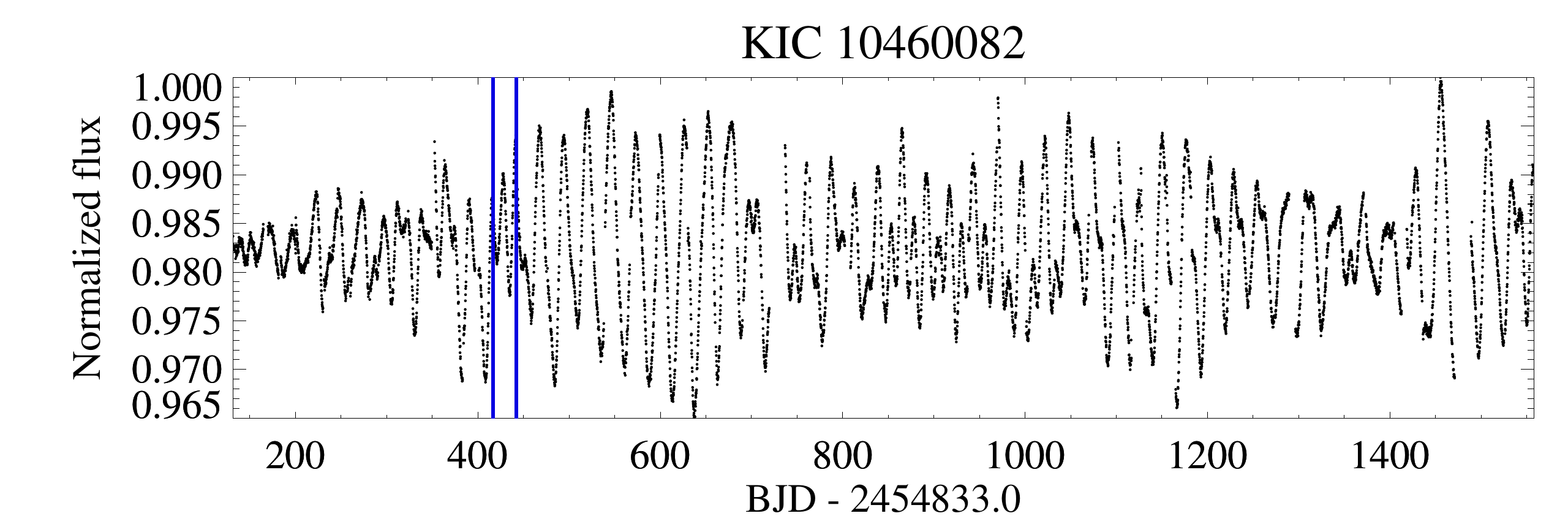}
	\includegraphics[scale=0.295]{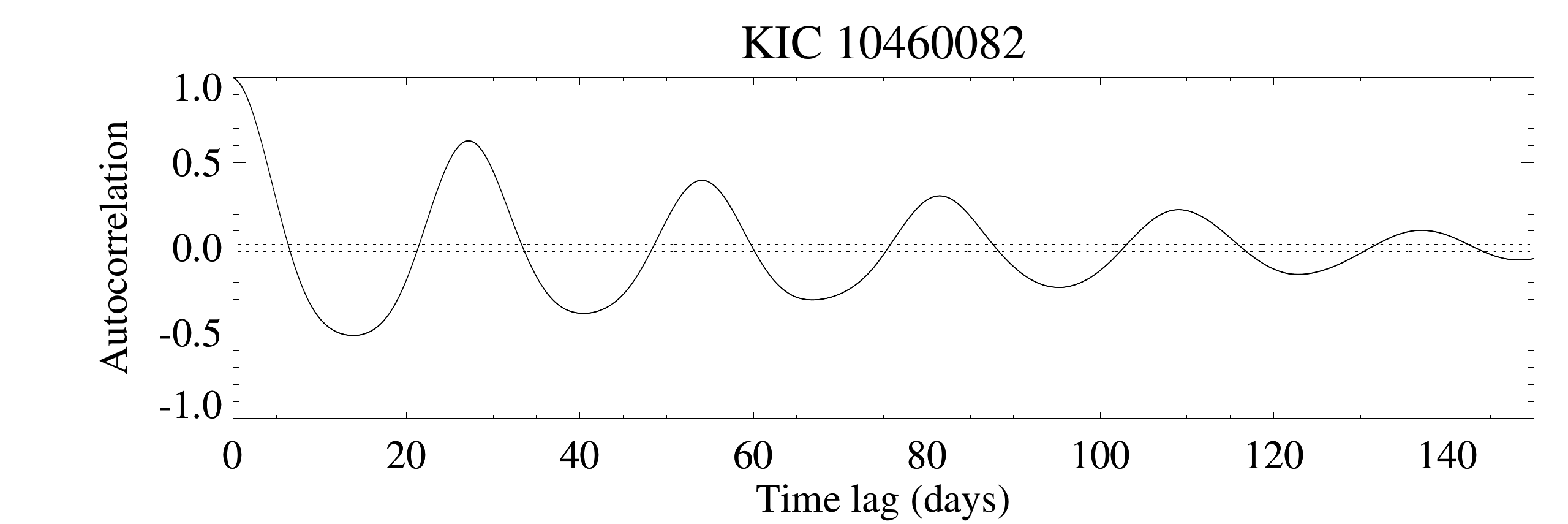}
	\includegraphics[scale=0.295]{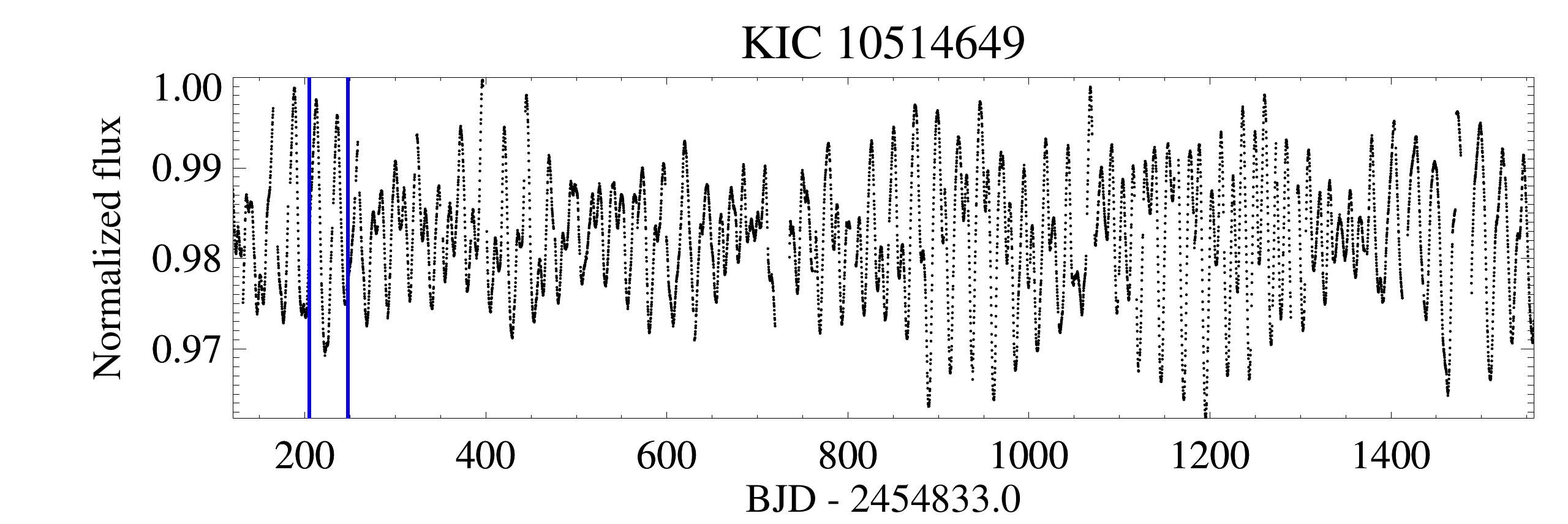}
	\includegraphics[scale=0.295]{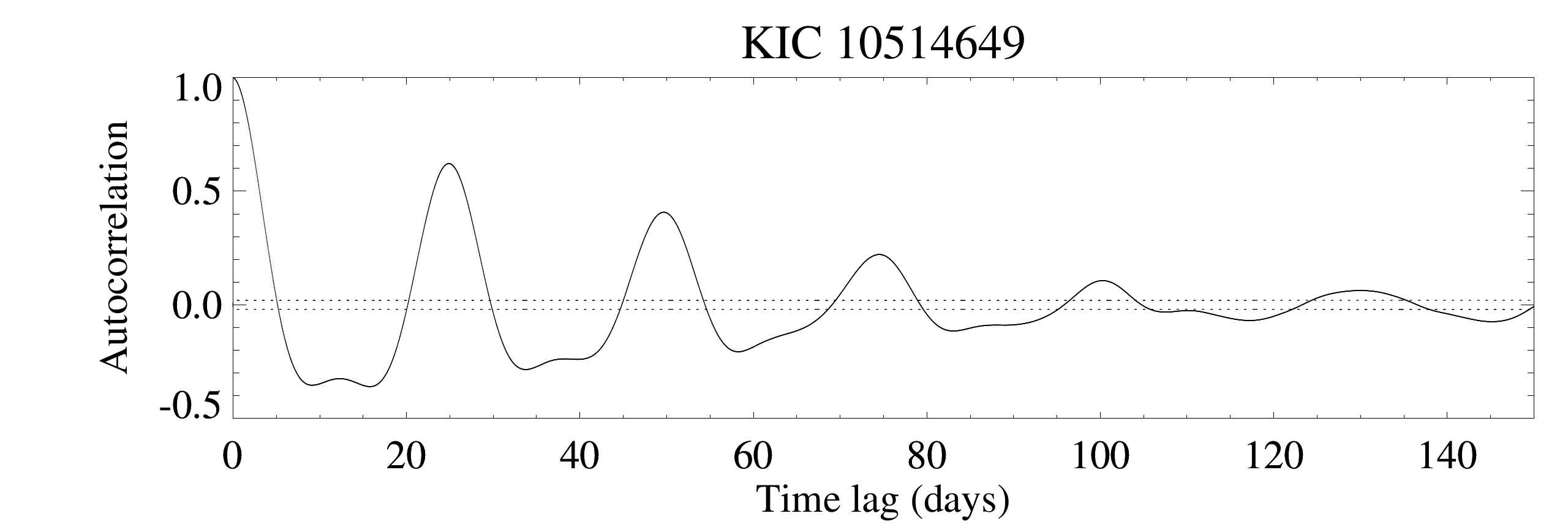}
	\includegraphics[scale=0.295]{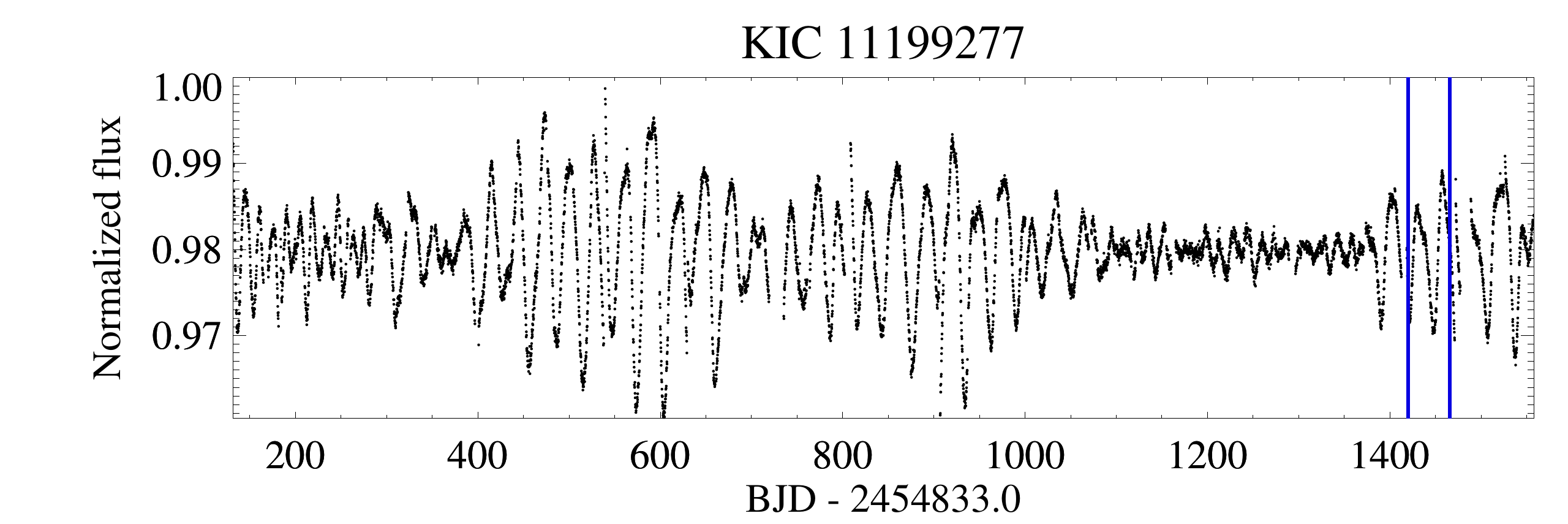}
	\includegraphics[scale=0.295]{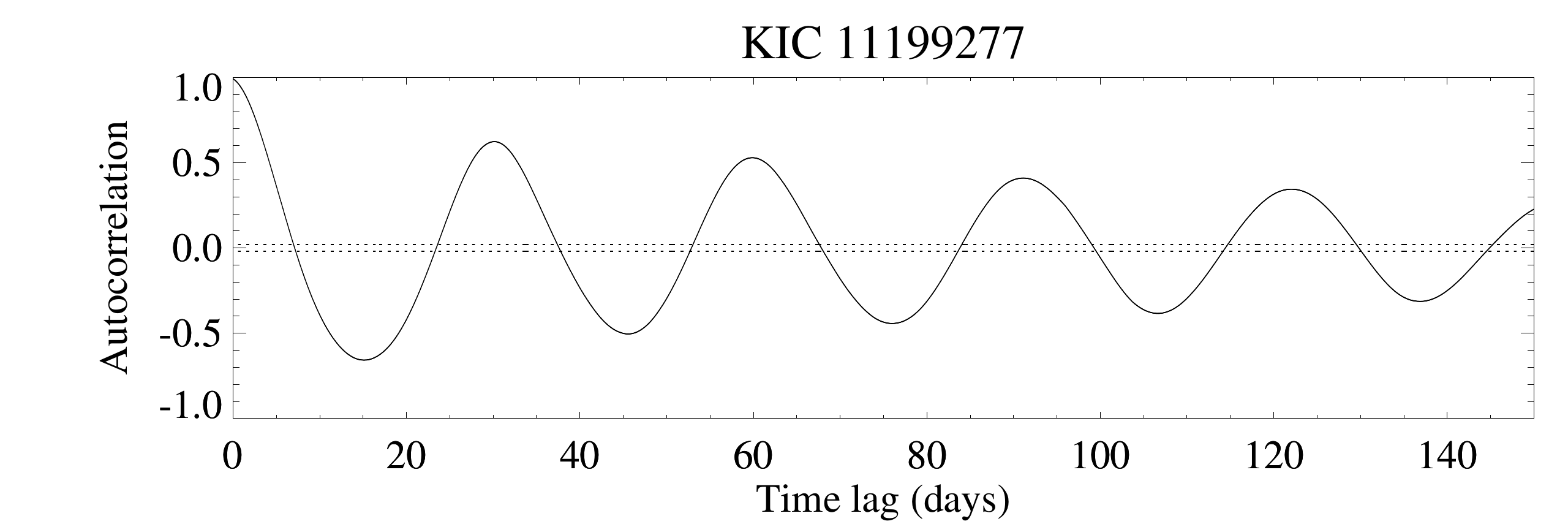}
	\includegraphics[scale=0.295]{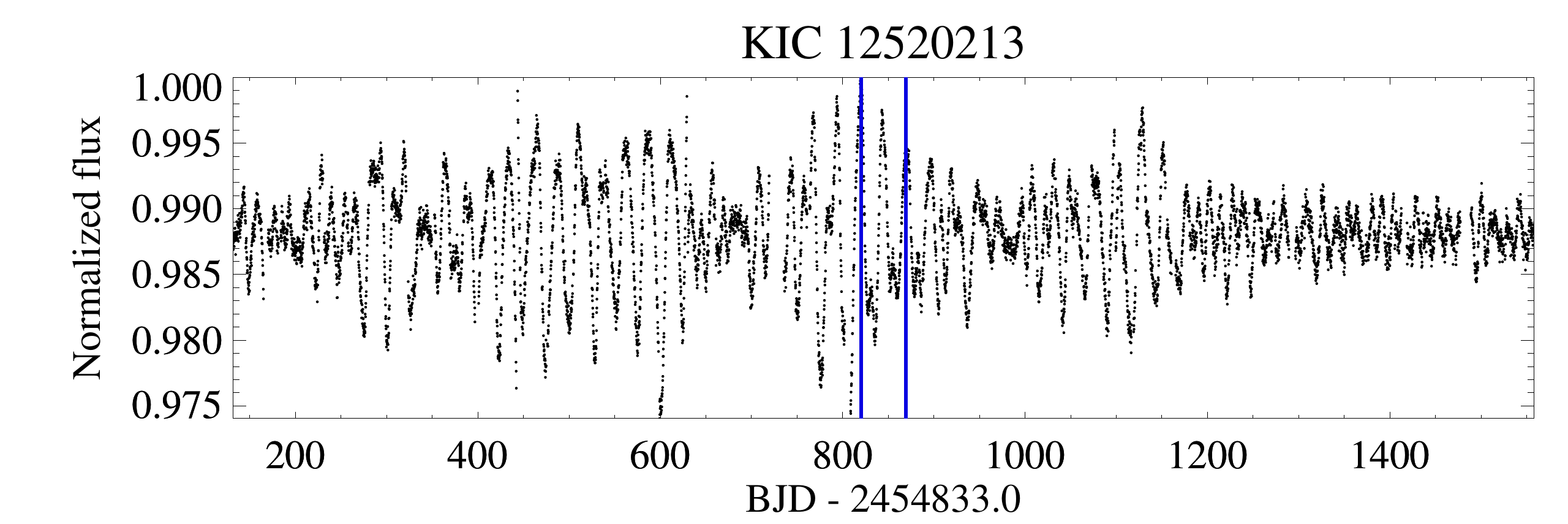}
	\includegraphics[scale=0.295]{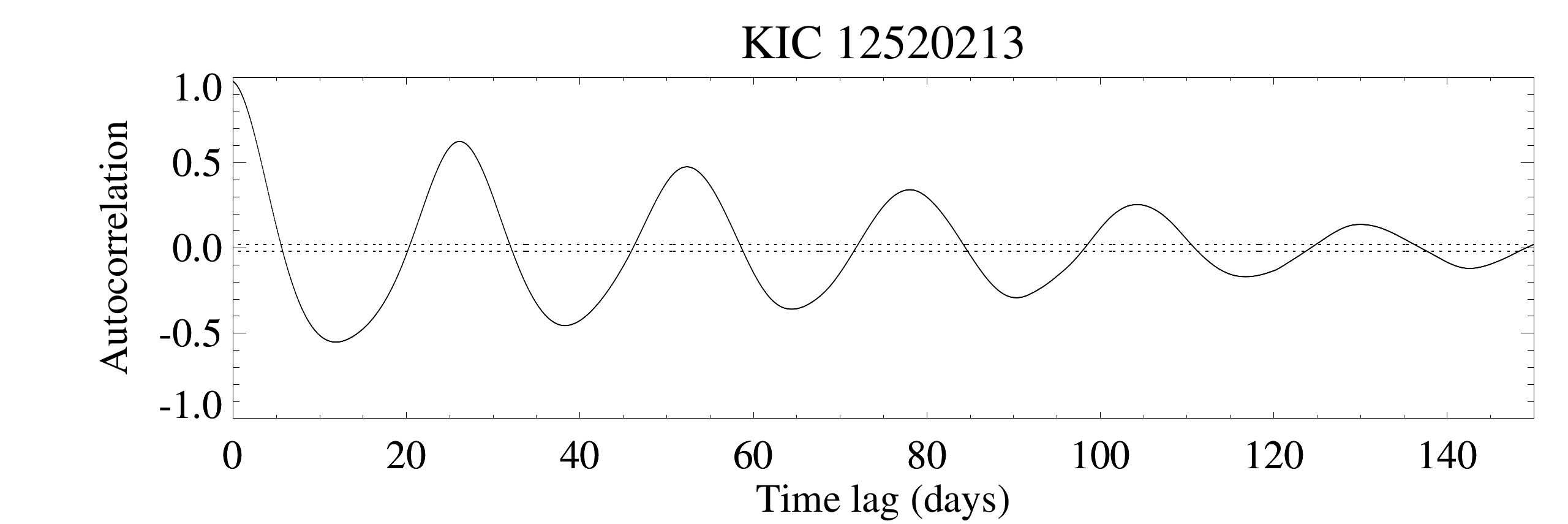}
	\caption*{.continued.}
	\label{FigLCAF2}%
\end{figure*}

\end{document}